\documentclass[11pt]{article}
\usepackage{times}
\usepackage{a4wide}

\usepackage{latexsym}
\usepackage{amsmath}
\usepackage{amssymb}
\usepackage{ifthen}
\usepackage{mathrsfs}
\usepackage{graphics}
\usepackage{color}
\usepackage{tikz}
\usepackage{stmaryrd}
\usepackage{amsthm}

\newcommand{\nc}{\newcommand}
\nc{\rnc}{\renewcommand} \nc{\nev}{\newenvironment}
\rnc{\subsection}{\secdef\ssa\ssb}
\nc{\ssa}[2][default]{\par\vspace{1ex}\refstepcounter{subsection}\noindent\textbf{\thesubsection.
#1. }} \nc{\ssb}[1]{\par\vspace{2ex}\noindent\textbf{#1. }}

\rnc{\subsubsection}{\secdef\sssa\sssb}
\nc{\sssa}[2][default]{\par\vspace{1ex}\refstepcounter{subsubsection}\noindent\textit{\thesubsubsection.
#1. }} \nc{\sssb}[1]{\par\vspace{1ex}\noindent\textit{#1. }}

\makeatletter
\rnc{\@seccntformat}[1]{{\normalfont\bfseries{\csname
the#1\endcsname}\hspace{1pt}.\hspace{0.4em}}}
\rnc{\section}{\@startsection
        {section}%
        {1}%
        {0mm}%
        {-\baselineskip}%
        {0.5\baselineskip}%
        {\normalfont\normalsize\bfseries\centering}%
}
\renewcommand{\@makecaption}[2]{\begin{center}#1. #2\end{center}}
\makeatother

\newtheorem{theo}{Theorem}[section]
\newtheorem{lem}[theo]{Lemma}
\newtheorem{cor}[theo]{Corollary}
\newtheorem{prop}[theo]{Proposition}

\theoremstyle{definition}
\newtheorem{defn}[theo]{Definition}

\rnc{\proof}[1][{}]{\smallskip\noindent\textit{Proof #1: }}
\nc{\proofend}{\hfill$\Box$\vspace{\topsep}\par}

\rnc{\labelenumi}{(\arabic{enumi})} \rnc{\labelitemi}{\text{--}}
\rnc{\phi}{\varphi} \rnc{\epsilon}{\varepsilon}
\nc{\bigmid}{\;\big|\;} \nc{\Bigmid}{\;\Big|\;}
\rnc{\max}{\textup{max}} \rnc{\min}{\textup{min}}
\rnc{\log}{\textup{log}\;}

\newlength{\probwidth}
\nc{\prob}[3][9]{
\begin{center}
  \normalfont\fbox{
   \begin{tabular}[t]{
     rp{#1cm}}\textit{Instance:}&#2. \\
     \textit{Problem:}&#3
   \end{tabular}}
\end{center}}

\nc{\pprob}[4][9]{
\begin{center}
   \normalfont\fbox{
    \begin{tabular}[t]{
     rp{#1cm}}\textit{Instance:}&#2. \\
     \textit{Parameter:}&#3. \\
     \textit{Problem:}&#4
   \end{tabular}}
\end{center}}

\nc{\nprob}[4][9]{
\begin{center}
  \normalfont\fbox{

\addtolength{\probwidth}{#1cm}\parbox{\probwidth}{\textsc{#2}\\\hspace*{1.5em}
     \begin{tabular}[t]{
      rp{#1cm}}\textit{Instance:}&#3. \\
      \textit{Problem:}&#4
     \end{tabular}}}
\end{center}}

\nc{\npprob}[5][9]{
\begin{center}
  \normalfont\fbox{

\addtolength{\probwidth}{#1cm}\parbox{\probwidth}{\textsc{#2}\\\hspace*{1.5em}
    \begin{tabular}[t]{
     rp{#1cm}}\textit{Input:}&#3. \\
     \textit{Parameter:}&#4. \\
     \textit{Question:}&#5
    \end{tabular}}}
\end{center}}

\nc{\nppxrob}[5][9]{ \normalfont\fbox{

\addtolength{\probwidth}{#1cm}\parbox{\probwidth}{\textsc{#2}\\\hspace*{1.5em}
   \begin{tabular}[t]{
    rp{#1cm}}\textit{Instance:}&#3. \\
    \textit{Parameter:}&#4. \\
    \textit{Problem:}&#5
   \end{tabular}}}}

\nc{\nppprob}[5][4]{
\begin{center}
  \normalfont\fbox{

\addtolength{\probwidth}{#1cm}\parbox{\probwidth}{\textsc{#2}\\\hspace*{1.5em}
    \begin{tabular}[t]{
     rp{#1cm}}\textit{Instance:}&#3. \\
     \textit{Parameter:}&#4. \\
     \textit{Problem:}&#5
    \end{tabular}}}
\end{center}}

\nc{\noptprob}[6][9]{
\begin{center}
  \normalfont\fbox{

\addtolength{\probwidth}{#1cm}\parbox{\probwidth}{\textsc{#2}\\\hspace*{1.5em}
    \begin{tabular}[t]{
     rp{#1cm}}\textit{Instance:}&#3. \\
     \textit{Solution:}&#4. \\
     \textit{Cost:}&#5. \\
     \textit{Goal:}&#6.
    \end{tabular}}}
\end{center}}

\nc{\FOR}{\textbf{for}}
\nc{\FORALL}{\textbf{for all}}
\nc{\TO}{\textbf{to}}
\nc{\DO}{\textbf{do}}
\nc{\OD}{\textbf{od}}
\nc{\IF}{\textbf{if}}
\nc{\FI}{\textbf{fi}}
\nc{\THEN}{\textbf{then}}
\nc{\ELSE}{\textbf{else}}
\nc{\WHILE}{\textbf{while}}
\nc{\REPEAT}{\textbf{repeat}}
\nc{\UNTIL}{\textbf{until}}
\nc{\OR}{\textbf{or}}
\nc{\AND}{\textbf{and}}
\nc{\PRINT}{\textbf{print}}

\nc{\im}[1]{\item\hspace{#1cm}}
\nev{algorithm}{\begin{enumerate}\rnc{\labelenumi}{\textit{\small \arabic{enumi}.}}\rnc{\itemsep}{0ex}}{\end{enumerate}}

\nc{\fpcl}[1]{\left[#1\right]_{\text{\upshape fp}}}
\nc{\pr}{\le^{\text{\normalfont fp}}_m} \nc{\FPT}{\textup{FPT}}
\nc{\EPT}{\textup{EPT}} \nc{\SUBEPT}{\textup{SUBEPT}}

\nc{\fpt}{\textup{fpt}} \nc{\fptT}{\textup{fpt-T}}
\nc{\W}[1]{\text{$\textup{W}[#1]$}}
\nc{\M}[1]{\text{$\textup{M}[#1]$}}
\nc{\MS}[2]{\text{$\textup{M}^{#1}[#2]$}}
\nc{\MINI}[1]{\mbox{\small \rm MINI[$#1$]}}
\nc{\WP}{\textup{W[P]}}
\nc{\AWP}{\textup{AW[P]}}
\nc{\AW}[1]{\text{$\textup{AW}[#1]$}}

\rnc{\S}[1]{\text{$\textup{S}[#1]$}} \nc{\SP}{\textup{S[P]}}
\nc{\MP}{\textup{M[P]}}

\nc{\PTIME}{\textup{PTIME}} \nc{\APTIME}{\textup{APTIME}}
\nc{\PSPACE}{\textup{PSPACE}} \nc{\NP}{\textup{NP}}

\nc{\DTIME}{\textup{DTIME}}

\nc{\se}{\subseteq} \nc{\re}{\rightarrow}

\nc{\LOEFF}[1]{{o}^{\rm eff}(#1)}

\nc{\PNPTC}{\mbox{$\textup{P}[{\textsc{tc}}]\ne
\textup{NP}[{\textsc{tc}}]$}} \nc{\str}[1]{\ensuremath{\mathcal #1}}
\nc{\cls}[1]{\ensuremath{\textup{#1}}}

\nc{\algo}[1]{{\mathbb #1}}

\nc{\NAT}{{\mathbb N}}

\nc{\VC}{\textsc{Vertex-Cover}}
\nc{\IS}{\textsc{Independent-Set}}
\nc{\Cli}{\textsc{Clique}}
\nc{\DS}{\textsc{Dominating-Set}}
\nc{\TSAT}{\textsc{3-Sat}}

\nc{\CNF}{\textup{CNF}} \nc{\WSAT}{\textsc{WSat}}
\nc{\AWSAT}{\textsc{AWSat}} \nc{\SAT}{\textsc{Sat}}
\nc{\ASAT}{\textsc{ASat}} \nc{\CIRC}{\textsc{Circ}}
\nc{\PROP}{\textsc{Prop}}

\nc{\nva}{\textup{nv}} \nc{\ncl}{\textup{nc}}

\nc{\ETH}{\textup{ETH}}
\nc{\Wone}{\textup{W[1]}}

\nc{\serf}{\textup{serf}} \nc{\serfT}{\textup{serf-T}}
\nc{\var}{\textup{var}}

\nc{\NUXP}{\textup{XP}_{\rm nu}} \nc{\XP}{\textup{XP}}
\nc{\SNP}{\textup{SNP}}
\nc{\X}{\textup{X}}

\nc{\ept}{{\rm ept}}

\nc{\E}{\textup{E}}

\nc{\HALT}{\textsc{Halt}}

\nc{\TM}{\textsc{TM}} \nc{\TMBA}{\textsc{TMBA}}

\nc{\ceil}[1]{\left\lceil#1\right\rceil}
\nc{\floor}[1]{\left\lfloor#1\right\rfloor}

\nc{\bende}{\eqno$\Box$} \nc{\benda}{\tag*{$\Box$}}

\nc{\pa}{\kappa}

\nc{\co}{\textup{co-}}

\rnc{\L}{\textup{LOGSPACE}}

\nc{\NL}{\textup{NLOGSPACE}}

\rnc{\P}{\textup{P}}

\rnc{\angle}[1]{\langle #1\rangle}

\nc{\rand}[1]{\marginpar{\raggedright\footnotesize #1}}
\nc{\yrand}[1]{\rand{\textbf{Y: }#1}}
\nc{\jrand}[1]{\rand{\textbf{J: }#1}}
\nc{\xrand}[1]{\rand{\textbf{X: }#1}}
\nc{\rjrand}[1]{\rand{\textcolor{red}{\textbf{J: }#1}}}

\nc{\Ppoly}{\textup{P}/\text{\small poly}}

\nc{\f}{\mathbf f}

\nc{\s}{\mathbf s}

\nc{\tim}{\textup{time}}

\nc{\sat}{\textup{sat}}
\nc{\rank}{\textup{rank}}
\nc{\PC}{\mathbf{PC}}
\nc{\bin}{{\rm in}}
\nc{\bout}{{\rm out}}
\nc{\Mix}{\textsc{Mix}}
\nc{\MIX}{\mathbf{MIX}}

\nc{\tdeg}{\textup{tdeg}}
\nc{\lspan}{\textup{span}}

\nc{\ma}[1]{\mathbb #1}
\nc{\bet}[1]{\| #1\|}
\nc{\EFM}[2]{(#1,#2)}
\nc{\EF}{\text{Ehren\-feucht\--Fra\"i\-ss\'e}}
\nc{\AF}{\textup{AF}}
\nc{\supp}{\textup{supp}}
\nc{\ar}{\textup{ar}}
\nc{\pol}{\textup{pol-}}
\nc{\equivm}{\equiv_{\LFP_m}}
\nc{\equivfo}{\equiv_{\FO_m}}
\nc{\ERRE}{Erd\H{o}s-R\'enyi}

\nc{\PCC}{\textup{PCC}}

\nc{\TCOL}{\textsc{3-Col}}

\nc{\GI}{\textup{GI}}

\nc{\EX}{\textup{E}}

\nc{\Var}{\textup{Var}}
\nc{\qr}{\textup{qr}}

\newcommand{\AC}{\textup{AC}}

\nc{\size}{\textup{size}}

\nc{\DNF}{\textup{DNF}}

\nc{\n}{\tilde n}

\nc{\dotcup}{\;\dot\cup\;}

\nc{\ds}{\gamma}

\nc{\mds}{\ensuremath{\textsc{Min-Dominating-Set}}}
\nc{\pds}{\ensuremath{p\textsc{-Dominating-Set}}}
\nc{\pClique}{\ensuremath{p\textsc{-Clique}}}
\nc{\pVC}{\ensuremath{p\textsc{-Vertex-Cover}}}
\nc{\pdHS}{\ensuremath{p\textsc{-$d$-Hitting-Set}}}

\nc{\pwsat}{\ensuremath{p\textsc{-WSat}}}

\nc{\mcs}{\ensuremath{\textsc{Monotone-Circuit-Satisfiability}}}

\nc{\vc}{\textsc{Vertex-Cover}}
\nc{\clique}{\textsc{Clique}}

\nc{\sol}{\textup{sol}}
\nc{\cost}{\textup{cost}}
\nc{\goal}{\textup{goal}}

\nc{\pow}{\text{Pow}}

\nc{\mwds}{\textsc{Min-Weighted-Dominating-Set}}

\nc{\paraAC}{\textup{para-$\AC^0$}}

\nc{\para}{\textup{para-}}

\nc{\stconn}{\textsc{stConn}}

\nc{\pstconn}{\ensuremath{p\textsc{-stConn}}}

\nc{\parity}{\textsc{Parity}}

\nc{\C}{\mathsf C}
\nc{\D}{\mathsf D}
\rnc{\PC}{\mathsf{PC}}

\nc{\FO}{\textup{FO}}

\nc{\phalt}{\ensuremath{p\textsc{-Halt}}}

\nc{\NTM}{\textup{NTM}}

\nc{\cn}{\ensuremath{\omega}}

\nc{\DTdepth}{\textup{DTdepth}}
\nc{\DTdepthv}{\ensuremath{\DTdepth_{\rm vertex}}}

\nc{\pgapclique}[1]{\ensuremath{p\textsc{-Gap}_{#1}\textsc{-Clique}}}
\nc{\pgapwsat}[1]{\ensuremath{p\textsc{-Gap}_{#1}\textsc{-Wsat}}}

\rnc{\log}{\textup{log}}

\nc{\res}[1]{\ensuremath{\!\!\upharpoonright_{#1}}}

\nc{\dlogtime}{\textup{dlogtime}}

\nc{\pacr}{\textrm{para-$\AC^0$}}
\nc{\pac}{\textrm{pac}}
\nc{\pwac}{\textrm{pwac}}

\nc{\LFP}{\textup{LFP}}

\nc{\G}{\mathbf G}

\nc{\uni}[1]{\ensuremath{[#1]}}
\nc{\ARITHM}{\textup{ARITHM}}
\nc{\pre}{\textit{pre}}
\nc{\la}{\langle}
\nc{\ra}{\rangle}
\nc{\Str}{\textit{Str}}
\nc{\red}{\textit{red}}
\nc{\td}{\textup{td}}

\nc{\Mod}{\textup{Mod}}

\rnc{\qr}{\textup{qr}}

\nc{\sipser}{\textsc{Sipser}}
\nc{\sipserf}{\textrm{Sipser}}

\nc{\enc}{\textup{enc}}

\nc{\arity}{\textup{arity}}

\nc{\true}{\textsc{true}}

\nc{\FD}{\textup{FD}}

\begin{document}

\title{Slicewise definability in first-order logic with bounded quantifier rank}

\author{Yijia Chen\\\normalsize School of Computer Science\\
\normalsize Fudan University\\
\normalsize yijiachen@fudan.edu.cn\\
\and
J\"{o}rg Flum\\\normalsize Mathematisches Institut \\
\normalsize Universit\"{a}t Freiburg\\
\normalsize joerg.flum@math.uni-freiburg.de \\
\and
Xuangui Huang\\\normalsize Department of Computer Science\\
\normalsize Shanghai Jiao Tong University\\
\normalsize stslxg@gmail.com}

\date{}
\maketitle

\begin{abstract}
For every $q\in \mathbb N$ let $\FO_q$ denote the class of sentences of
first-order logic \FO\ of quantifier rank at most $q$. If a graph property
can be defined in $\FO_q$, then it can be decided in time $O(n^q)$. Thus,
minimizing $q$ has favorable algorithmic consequences. Many graph
properties amount to the existence of a certain set of vertices of size
$k$. Usually this can only be expressed by a sentence of quantifier rank
at least $k$. We use the color-coding method to demonstrate that some
(hyper)graph problems can be defined in $\FO_q$ where $q$ is independent
of $k$. This property of a graph problem is equivalent to the question of
whether the corresponding parameterized problem is in the class
$\para\AC^0$.

It is crucial for our results that the \FO-sentences have access to
built-in addition and multiplication. It is known that then \FO\
corresponds to the circuit complexity class uniform $\AC^0$. We explore
the connection between the quantifier rank of \FO-sentences and the depth
of $\AC^0$-circuits, and prove that $\FO_q \subsetneq \FO_{q+1}$ for
structures with built-in addition and multiplication.
\end{abstract}

\subsection*{Keywords} first-order logic, quantifier rank, parameterized $\AC^0$, circuit depth.

\section{Introduction}
Let $\varphi$ be a sentence of first-order logic \FO. The \emph{quantifier
rank} of $\varphi$, denoted by $\qr(\varphi)$, is the maximum nested depth
of quantifiers in $\varphi$. If $\varphi$ defines a graph property $\cls K$,
that is,
\[
\cls K= \big\{\str G \bigmid \text{$\str G$ a graph and $\str G$ has the
property $\varphi$}\big\},
\]
then a straightforward algorithm can decide whether an input graph $\str G$
belongs to $\cls K$ in time $O(|\str G|^{\qr(\varphi)})$.
Therefore, minimizing the quantifier rank of $\varphi$ would lead to better
algorithms for deciding the graph property $\cls K$. Many graph properties
amount to the existence of a certain set of vertices of size $k$, where $k$
is a fixed constant. A well-known example is the \emph{$k$-vertex-cover
problem} of deciding whether a given graph $\str G$ contains a set $C$ of
$k$ vertices such that every edge in $\str G$ has one end in $C$.
%
%
The set $C$ is then called a \emph{$k$-vertex-cover} of $\str G$. Clearly,
the existence of a $k$-vertex-cover can be expressed by the following
sentence of \FO\
\[
\psi_k:= \exists x_1 \cdots \exists x_k
       \left(\bigwedge_{1\le i< j\le k} x_i\ne x_j \wedge
        \forall u \forall v \big(Euv \to \bigvee^k_{i=1} (u= x_i \vee v=x_i)\big)\right).
\]
In other words, a graph $\str G$ has a $k$-vertex-cover if and only if $\str
G$ satisfies $\psi_k$. Observe that $\qr(\psi_k)=k+2$, hence the naive
algorithm derived from $\psi_k$ would have running time $O(|\str G|^{k+2})$.
Clearly it is far worse than the existing linear time algorithms for the
$k$-vertex-cover problem. An immediate question is whether the
$k$-vertex-cover problem can be defined by a sentence $\varphi_k$ with
$\qr(\varphi_k)< k+2$. As the first main result of this paper we show that
this is in indeed possible for a $\varphi_k$ with $\qr(\varphi_k)\le 16$.
Note that this holds for every $k$ even though we need different
$\varphi_k$'s for different $k$'s. The $k$-vertex-cover problem is the
\emph{$k$th slice} of the parameterized vertex cover problem
\npprob[7.5]{\pVC}{A graph $\str G$}{$k$}{Does $\str G$ have a vertex cover
of size $k$?}
For $q\in\mathbb N$ we denote by $\FO_q$ the class of \FO-sentences of
quantifier rank at most $q$. Our result can be phrased in terms of the
\emph{slicewise definability}~\cite{flugro03} of $\pVC$:
\begin{theo}\label{thm:pvc}
$\pVC$ is slicewise definable in $\FO_{16}$.
\end{theo}

The vertex cover problem is a special case of the \emph{hitting set problem}
on hypergraphs of bounded hyperedge size. For every $d\in \mathbb N$ a
$d$-hypergraph is a hypergraph with hyperedges of size at most $d$. Then,
the \emph{parameterized $d$-hitting set problem} $\pdHS$ asks whether an
input $d$-hypergraph $\str G$  contains a set of $k$ vertices that
intersects with every hyperedge in~$\str G$. 
Thus $\pVC$ is basically the parameterized $2$-hitting set problem.
Extending Theorem~\ref{thm:pvc} we prove that $\pdHS$ is slicewise definable
in $\FO_{q}$, where $q= O(d^2)$. The problem $\pdHS$ can be
Fagin-defined~\cite{flugro01} by an \FO-formula with a second-order variable
which does not occur in the scope of an existential quantifier or negation
symbol. We show that all problems Fagin-definable in this form are slicewise
definable in some $\FO_q$.

What is the complexity of the class of parameterized problems that are
slicewise definable in \FO\ with bounded quantifier rank? We prove that it
coincides with $\para\FO$~\cite{cheflu16}, the class of problems
\FO-definable after a precomputation on the parameter. Thus we obtain a
descriptive characterization of the class $\para\FO$, or equivalently of the
parameterized circuit complexity class
$\para\AC^0$~\cite{elbstotan15,banstotan15,cheflu16}.

The equivalence between $\para\FO$ and $\para\AC^0$ is an easy consequence
of the equivalence between \FO\ and the classical circuit complexity class
uniform $\AC^0$~\cite{barimm90}. This equivalence crucially relies on the
assumption that the input graphs (or more generally, the input structures)
are equipped with built-in addition and multiplication. In fact, the main
technical tool for proving Theorem~\ref{thm:pvc} and the subsequent results,
the \emph{color-coding} method~\cite{aloyus95}, makes essential use of
arithmetic. Without addition and multiplication, it is not difficult to show
that $\pVC$ cannot be slicewise defined in $\FO_q$ for any $q\in \mathbb N$.
Thus Theorem~\ref{thm:pvc} exhibits the power of addition and
multiplication, although on the face of it, the vertex cover problem has
nothing to do with arithmetic operations.

In finite model theory there is consensus that inexpressibility results for
\FO\ and for fragments of \FO\ are very hard to obtain in the presence of
addition and multiplication. To get such a result we exploit the equivalence
between \FO\ and uniform $\AC^0$, more precisely, we analyze the connection
between the quantifier rank of a sentence $\varphi$ and the depth of the
corresponding $\AC^0$ circuits. Together with a
theorem~\cite{has89,segbusimp04} on a version of Sipser functions we show
that the hierarchy $(\FO_q)_{q\in \mathbb N}$ is strict:

\begin{theo}\label{thm:FOqhier}
Let $q\in \mathbb N$. Then there is a parameterized problem slicewise
definable in $\FO_{q+1}$ but not in $\FO_q$.
\end{theo}

\subsection*{Organization of the paper}
In Section~\ref{sec:pvc} we prove Theorem~\ref{thm:pvc}, and then extend it
to the hitting set problem in Section~\ref{sec:phs}. We give a natural class
of Fagin-definable problems that are slicewise definable in \FO\ with
bounded quantifier rank in Section~\ref{sec:fagin}. We prove the hierarchy
theorem, i.e., Theorem~\ref{thm:FOqhier}, in Section~\ref{sec:FOqhier}. In
the final section we conclude with some open problems.

\subsection*{Some logic preliminaries}
A \emph{vocabulary} $\tau$ is a finite set of relation symbols. Each
relation symbol has an \emph{arity}. A \emph{structure}~$\str{A}$ of
vocabulary $\tau$, or \emph{$\tau$-structure}, con\-sists of a nonempty
set~$A$ called the \emph{universe} of $\str A$, and of an interpretation
$R^{\str{A}}\subseteq A^r$ of each $r$-ary relation symbol $R\in \tau$. In
this paper all structures have a finite universe. Occasionally we allow the
use of constants: For a vocabulary $\tau$ we consider $\tau\cup
\{c_1,\ldots, c_s\}$-structures $\str A$. Then $c_1^{\str A},\ldots,
c_s^{\str A}$, the interpretations of the constants $c_1,\ldots, c_s $, are
elements of $\str A$. However the letters $\tau$, $\tau'$, \ldots will
always denote \emph{relational} vocabularies (without constants). If $\tau$
contains a binary relation symbol $<$ and in the structure $\str A$ the
relation $<^{\str A}$ is an order of the universe, then $\str A$ is an
\emph{ordered structure}.

Let $\tau$ be a vocabulary and $C$ a set of constant. Formulas $\varphi$ of
first-order logic of vocabulary $\tau\cup C$ are built up from atomic
formulas $t_1=t_2$ and $Rt_1 \ldots t_r$ where $t_1,t_2,\ldots, t_r$ are
either variables or constants in $C$, and where $R\in\tau$ is of arity $r$,
using the Boolean connectives and existential and universal quantification.
A formula $\varphi$ is a \emph{sentence} if it has no free variables. The
quantifier rank of $\varphi$ is defined inductively as:
\begin{align*}
\qr(\varphi) :=
\begin{cases}
0 & \text{if $\varphi$ is atomic} \\
\qr(\psi) & \text{if $\varphi= \neg \psi$} \\
\max\{\qr(\psi_1), \qr(\psi_2)\} & \text{if $\varphi= \psi_1\wedge \psi_2$ or $\varphi= \psi_1\vee \psi_2$} \\
1+\qr(\psi)&\text{if $\varphi= \exists x\psi$ or $\varphi= \forall x \psi$}.
\end{cases}
\end{align*}

\section{Slicewise-definability in $\FO_q$ and the vertex cover problem}\label{sec:pvc}
In this section we prove Theorem~\ref{thm:pvc}, i.e., $\pVC$ is slicewise
definable in $\FO_{16}$. Our main tool is Theorem~\ref{thm:ccc}. It shows
how we can express that there are $k$ elements having a first-order property
by a number of quantifiers independent of $k$. We give further applications
of this tool in this and the next section.

For $n\in \mathbb N$ let $\uni n:= \{0,1,\ldots, n-1\}$. Denote by $<^{\uni
n}$ the natural order on $\uni n$. Clearly, if~$\str A$ is any ordered
structure, then $\big(A, <^{\str A}\big)$ is isomorphic to $\left([|A|],
<^{[|A|]}\right)$ and the isomorphism is unique. For ternary relation
symbols $+$ and $\times$ we consider the ternary relations $+^{\uni n}$ and
$\times^{\uni n}$ on~$\uni n$ that are the relations of addition and
multiplication of~$\mathbb N$ restricted to $\uni n$. That is,
\begin{equation*}
\begin{array}{rl}
+^{\uni n} &:= \big\{(a,b,c)\bigmid \text{$a,b,c\in \uni n$ with $c=a+b$}\big\}, \\[1mm]
\times^{\uni n} &:= \big\{(a,b,c)\bigmid \text{$a,b,c\in \uni n$ with $c=a\cdot b$}\big\}.
\end{array}
\end{equation*}
Finally, for every $m\in \mathbb N$ let $C(m):= \big\{\overline \ell \bigmid
\ell< m\big\}$ be a set of constants and set
\[
\overline \ell\,^{\uni n} :=\ell, \ \ \text{if $\ell<n$}
 \qquad \text{and}\qquad \overline \ell\,^{\uni n} :=n-1, \ \ \text{if $\ell\ge n$}.
\]
Assume a relational vocabulary $\tau$ contains $<$, $+$, and $\times$. A
$\tau\cup C(m)$-structure $\str A$ \emph{has built-in $<$ ,$+$, $\times$,
$C(m)$} if its $\{<, +, \times, C(m)\}$-reduct is isomorphic to $\left(\uni
n, <^{\uni n}, +^{\uni n}, \times^{\uni n}, (\overline \ell\,^{\uni
n})_{\ell< m}\right)$.

If $m=0$, we briefly say that $\str A$ \emph{has built-in addition and
multiplication}. We denote by $\ARITHM[\tau]$ the class of $\tau$-structures
with built-in addition and multiplication. If $\str A\in \ARITHM[\tau]$ and
$m\in\mathbb N$, we denote by $\str A_{C(m)}$ its unique expansion to a
$\tau\cup C(m) $-structure with built-in $<, +, \times, C(m)$.

\medskip
In the proof of Theorem~\ref{thm:ccc} we use the color-coding technique of
Alon et al.~\cite{aloyus95} essentially in the form presented in~\cite[page
347]{flugro06}:

\begin{lem}\label{lem:ccc}
There is an $n_0\in\mathbb N$ such that for all $n\ge n_0$, all $k\le n$ and
for every $k$-element subset $X$ of $[n]$, there exists a prime $p <
k^2\cdot \log_2\; n$ and a $q< p$ such that the function $h_{p,q}: [n]\to
\{0, \ldots, k^2 -1\}$ given by $h_{p,q}(m):= (q\cdot m \mod p) \mod k^2$ is
injective on $X$.
\end{lem}

As already mentioned the following result allows to express the existence of
$k$ elements satisfying a first-order property by a bounded number of
quantifiers.

\begin{theo}\label{thm:ccc}
Let $\tau$ be a vocabulary containing $<$, $+$, $\times$. Then there is an
algorithm that assigns to every $k\in \mathbb N$ and every
$\FO[\tau]$-formula $\varphi(\bar x, y)$ an $\FO\big[\tau\cup
C(k^2+1)\big]$-formula $\chi_{\varphi}^k(\bar x)$ such that for every $\str
A\in \ARITHM[\tau]$ with $k^2\le |A|/\log\; |A|$ and $|A|\ge n_{0}$ and
$\bar u\in A$,
\begin{equation}\label{eq:ccvarphi}
\begin{array}{l}
\str A_{C(k^2)}
 \models \chi^k_{\varphi}(\bar u) \iff \\[1mm]
{\hspace{.5cm}} \text{there are pairwise distinct $v_0, \ldots, v_{k-1}\in A$ with
 $\str A\models \varphi(\bar u,v_i)$ for every $i\in [k]$}.
\end{array}
\end{equation}
Furthermore, $\qr\big(\chi^k_{\varphi}(\bar x)\big)= \max\big\{12,
\qr\big(\varphi(\bar x, y)\big)+3\big\}$.

Note that the conditions ``$k^2\le |A|/\log\; |A|$ and $|A|\ge n_{0}$'' on
$|A|$ are fulfilled if $|A|\ge \max\left\{2^{k^2}, n_{0}\right\}$, so we
have a lower bound of $|A|$ in terms of $k$ (here~$ n_0$ is a natural number
according to Lemma~\ref{lem:ccc}).
\end{theo}

\proof Let $\str A$ be as above, set $n:=|A|$, and w.l.o.g. assume that
$A:=\uni {n}$. In order to make formulas more readable, we introduce some
abbreviations. Clearly, $x = (y\!\! \mod z)$ is an abbreviation for
\[
\exists u (y= u\times z+x \wedge x < z),
\]
more precisely, as $+$ and $\times$ are relation symbols, an abbreviation
for
\[
\exists u\exists u'(u'=u\times z\wedge y= u'+x\wedge x < z).
\]
Now let
\[
\chi^k_{\varphi}(\bar x)
 := \exists p\exists q
  \Big(\bigvee_{0\le i_1<\ldots <i_{k-1}<k^2}
   \bigwedge_{j\in[k]}
    \exists y \big(\text{``$h_{p,q}(y)=i_j$''} \wedge \varphi(\bar x, y)\big)\Big),
\]
where
\[
\text{``$h_{p,q}(y)=i_j$''}
 :=(q\times (u \!\! \mod p) \!\!\mod p) \!\!\mod \overline{k^2}\
  = \overline{i_j}.
\]
We replaced $(q\times u\!\!\mod p)$ by $(q\times (u \!\! \mod p) \!\!\mod
p)$, since $q\times u$ might exceed $|A|$. To count the quantifier rank note
that ``$h_{p,q}(y)=i_j$'' means
\[
\exists v\exists v'\exists \alpha \left(v'=v\times \overline{k^2}\wedge
\alpha=v'+\overline{i_j}\wedge \overline{i_j}<\overline{k^2}\right),
\]
where the intended meaning of $\alpha$ is $(q\times (u \!\! \mod p) \!\!\mod
p)$. So $\alpha$ is the unique element satisfying
\[
\exists w\exists w'\exists \beta( w'=w\times p\wedge \beta=w'+\alpha\wedge \alpha <p).
\]
Here the intended meaning of $\beta$ is $q\times (u \!\! \mod p)$. Thus
$\beta$ is the unique element satisfying
\[
\exists \gamma(\beta=q\times \gamma\wedge \text{ ``$ \gamma=u\!\! \mod p$''}).
\]
So we can replace ``$\gamma=u\!\! \mod p$'' by
\[
\exists z\exists z'(z'=z\times p\wedge u=z'+\gamma \wedge \gamma <p).
\]
Thus, $\qr\big(\text{``$h_{p,q}(y)=i_j$''}\big)=9$ and hence,
$\qr\big(\chi^k_{\varphi}(\bar x)\big)= \max\big\{12, \qr\big(\varphi(\bar
x, y)\big)+3\big\}$. \proofend

We use the previous result to show that two parameterized problems are
slicewise definable in $\FO_{q}$ for some $q$, one is an easy application,
the other the more intricate $\pVC$. First we give the precise definitions
of parameterized problem in our context and of slicewise definability.

\begin{defn}\label{def:pap}
A parameterized problem is a subclass $Q$ of $\ARITHM[\tau]\times \mathbb N$
for some vocabulary~$\tau$, where for each $k\in\mathbb N$ the class $Q_k:=
\{\str A\in \ARITHM[\tau] \mid (\str A, k)\in Q\} $ is closed under
isomorphism. The class $Q_k$ is the \emph{$k$th slice of} $Q$.

Every pair $(\str A, k)\in \ARITHM[\tau]\times \mathbb N$ is an
\emph{instance} of $Q$, \str A its \emph{input} and $k$ its
\emph{parameter}.
\end{defn}

\begin{defn}\label{def:sli}
$Q$ is \emph{slicewise definable in \FO\ with bounded quantifier rank},
briefly $Q\in\X\FO_{\qr}$, if there is a $q\in \mathbb N$ and computable
functions $h:\mathbb N\to \mathbb N$ and $f:\mathbb N\to \mathbb
\FO_q[\tau\cup C(h(k))]$ such that for all $(\str A, k)\in
\ARITHM[\tau]\times \mathbb N$,
\[
(\str A, k)\in Q\iff \str A_{C(h(k))}\models f(k).
\]
That is, if $m_k:=h(k)$ and $f(k):=\varphi_k$, then
\[
(\str A, k)\in Q\iff \str A_{C(m_k)}\models \varphi_k.
\]
We then say that $Q$ \emph{is slicewise definable in $\FO_q$} and write
$Q\in \X\FO_q$.
\end{defn}

Using the constants in $C(m)$ we can characterize arithmetical structures
with less that $m$ elements by a quantifier free sentence, more precisely:
\begin{lem}\label{lem:charstr}
Assume that $\str A\in \ARITHM[\tau]$ and that $|A|< m$. Then there is a
quantifier free $\FO[\tau\cup C(m)]$-sentence $\varphi_{\str A_{C(m)}}$
\big(that is, $\varphi_{\str A_{C(m)}}\in \FO_0[\tau\cup C(m)]$\big) such
that for all structures $\str{B}\in \ARITHM[\tau]$ we have
\[
\str{B}_{C(m)}\models \varphi_{\str A_{C(m)}}\iff \str A\cong \str{B}.
\]
\end{lem}

Using this lemma we get the following simple but useful observation.
\begin{prop}\label{pro:evesli}
Let $Q\in \ARITHM[\tau]\times \mathbb N$ be a decidable parameterized
problem and $q\in\mathbb N$. Assume that $Q$ is \emph{eventually slicewise
definable in $\FO_q$}, that is, there are computable functions $k\mapsto
m_k$ with $m_k\in \mathbb N$ and $k\mapsto \varphi_k$ with $\varphi_k\in
\FO_q\big[\tau\cup C(m_k)\big]$ and a computable and increasing function $g:
\mathbb N\to \mathbb N$ such that for all $(\str A, k)\in
\ARITHM[\tau]\times \mathbb N$ with $|A|\ge g(k)$,
\[
(\str A, k)\in Q\iff \str A_{C(m_k)}\models \varphi_k.
\]
Then $Q$ is slicewise definable in $\FO_q $.
\end{prop}

\proof Assume $Q$ is eventually slicewise definable in $\FO_q$ and let
$m_k$, $\varphi_k$, and $g$ be  as above. The sentence $\psi_k$ defining the
$k$th slice of $Q$ essentially says
\begin{align*}
\text{\big(the structure has}
 & \text{ at least $g(k)$ elements and satisfies $\varphi_k$\big) or}\\
 & \text{\big(the structure has less than $g(k)$ elements and is in $Q$\big).}
\end{align*}
To express this we use the set $C(m_k')$ of constants where
$m'_k:=\max\{g(k),m_k \}$. In structures with built-in $<,+,\times$
and~$C(m'_k)$ the sentence $\overline{g(k)-1}\ne \overline{g(k)-2}$ says
that the universe has $\ge g(k)$ elements. So we can set (compare
Lemma~\ref{lem:charstr})
\[
\psi_k:= \Big( \overline{g(k)-1}\ne \overline{g(k)-2} \wedge \varphi_k\Big)
 \vee \bigvee_{(\str A, k)\in Q,\ |A|< g(k)}\varphi_{\str A_{C(g(k))}}.
\]
Hence, the quantifier rank of each $\psi_k$ coincides with the quantifier
rank of $\varphi_k$. As $Q$ is decidable, the mapping $k\mapsto \psi_k$ is
computable. \proofend

We now turn to our first application of Theorem~\ref{thm:ccc}.

\begin{theo}
The parameterized problem
\npprob[8]{$p\textit{-deg-}\textsc{Independent-Set}$}{A graph $\str
G$}{$k\in\mathbb N$}{Is $k\ge\textup{deg}(\str G)$ and does \str G have an
independent set of $k-\textup{deg}(\str G)$ elements?}
is slicewise definable in $\FO_{13}$.
\end{theo}
Let $\tau_{\textsc{Graph}}:=\{E,<,+,\times\}$ with binary $E$. More
formally, by
$p\textit{-deg-}\textsc{Independent-Set}$ we mean in our context  the  class
\begin{align*}
  \Big\{(\str G,k)& \in \ARITHM[\tau_{\textsc{Graph}}] \times \mathbb N
    \Bigmid k\ge\textup{deg}(\str G) \ \text{and} \\
 & \text{\big(the $\{E\}$-reduct of\big) $\str G$ has an independent set
   of size $ k-\textup{deg}(\str G)$} \Big\}.\footnotemark
\end{align*}
\footnotetext{In the following we will present parameterized graph problems
in the more liberal form as given by the box above.}
\proof
An easy induction on $\ell:= k-\textup{deg}(\str G)$ shows that every graph
$\str G$ with at least $(\textup{deg}(\str G)+1)\cdot \ell$ vertices has an
independent set of size $\ell$. Hence, for $(\str G,k)\in \ARITHM[\tau]$,
where the graph $\str G$ has at least $(k+1)\cdot k$ vertices, we have
\begin{eqnarray}\label{eq:kgde}
(\str G,k)\in p\textit{-deg-}\textsc{Independent-Set}
 & \iff &
 k\ge \textup{deg}(\str G).
\end{eqnarray}
We use this fact to prove that $p\textit{-deg-}\textsc{Independent-Set}$ is
eventually slicewise definable in $\FO_{13}$, which yields our claim by
Proposition~\ref{pro:evesli}.

Let $d\in\mathbb N$ and $\varphi(u,y):= Euy$. Then, by
Theorem~\ref{thm:ccc}, we have for every graph $\str G$ with at least $h(k)$
vertices for some computable $h:\mathbb N\to \mathbb N$ and every vertex $u$
of $\str G$,
\begin{eqnarray*}
\str G\models\chi^d_{\varphi}(u)
 & \iff &
\text{the degree of $u$ in $\str G$ is $\ge d$}.
\end{eqnarray*}
So the degree of $\str G$ is the unique $d$ such that
\[
\str G\models \exists
u\chi^d_{\varphi}(u)\wedge \neg \exists u\chi^{d+1}_{\varphi}(u).
\]
Thus, for
$k\in\mathbb N$ and every graph $\str G\in \ARITHM[\tau_{\textsc{Graph}}]$
with at least $\max\big\{h(k), (k+1)\cdot k \big\}$ vertices,
by~\eqref{eq:kgde},
\begin{eqnarray*}
(\str G, k)\in p\textit{-deg-}\textsc{Independent-Set}
 & \iff &
\str G\models\bigvee_{d\le k}\left(\exists u\chi^d_{\varphi}(u)\wedge \neg \exists u\chi^{d+1}_{\varphi}(u)\right).
\end{eqnarray*}
As $\qr(\varphi)=0$, Theorem~\ref{thm:ccc} and the previous equivalence show
that $p\textit{-deg-}\textsc{Independent-Set}$ is eventually in $\X\FO_{13}$
\big(and hence in $\X\FO_{13}$ by Proposition~\ref{pro:evesli}\big).
\proofend

Now we are ready to show the slicewise definability of $\pVC$ in $\FO_{16}$.

\proof[of Theorem~\ref{thm:pvc}.] Recall the main ingredient of Buss'
kernelization for an instance $(\str G,k)$ of the vertex cover problem.
\begin{enumerate}
\item If a vertex $v$ has degree $\ge k+ 1$ in $\str G$, then $v$ must be
    in every vertex cover of size $k$. We remove all $v$ of degree $\ge k+
    1$ in $\str G$, say $\ell$ many, and decrease $k$ to $k':= k- \ell$.

\item Remove all isolated vertices.

\item Let $\str G'$ be the resulting induced graph. If $k'<0$ or $\str G'$
    has $> k'\cdot (k+1)$ vertices, then $(\str G',k')$, and hence also
    $(\str G,k)$, is a \textsc{no} instance of $\pVC$.
\end{enumerate}
Again let $\varphi(x,y):=Exy$. Then, by Theorem~\ref{thm:ccc}, for every
instance $(\str G, k)$ of $\pVC$, where the vertex set $G$ of $\str G$ is
sufficiently large compared with $k$ and every vertex $v\in G$,
\begin{eqnarray*}
\str G\models\chi_{\varphi}^{k+1}(v)
 & \iff &
\text{$v$ has degree $\ge k + 1$}.
\end{eqnarray*}
Therefore, applying again Theorem~\ref{thm:ccc} we get for $\ell\in \mathbb
N$,
\begin{eqnarray*}
\str G\models
 \left(\chi^{\ell}_{\chi_{\varphi}^{k+1}}\wedge \neg\chi^{\ell+1}_{\chi_{\varphi}^{k+1}}\right)
 & \iff &
\text{$\str G$ has exactly $\ell$ vertices of degree $\ge k + 1$}.
\end{eqnarray*}
For every vertex $v$ of $\str G$ we have
\begin{eqnarray*}
\str{G }\models \textit{uni}(v)
 & \iff &
\text{$v$ is a vertex of $\str G'$},
\end{eqnarray*}
where
\[
\textit{uni}(x):=
\left(\neg\chi_{\varphi}^{k+1}(x)\wedge
 \neg \forall y\big(Exy\to \chi_{\varphi}^{k+1}(y)\big)\right).
\]
Then,
\begin{align}\label{eq:tab}
 (\str G,&k)\in \pVC \notag \\[1mm]
  & \iff
    \text{for some $\ell$ with $0\le\ell \le k$,
          $\str G$ has exactly $\ell$ vertices of degree $\ge k+ 1$ and} \notag \\
  & \hspace*{3.2cm}
  \text{there is a $j\le (k-\ell)\cdot (k+1)$ such that $\str{G }'$ has $j$ vertices and}\notag \\
  & \hspace*{4.3cm}
  \text{$(\str G',k-\ell)$ is a \textsc{yes} instance of $\pVC$ }\notag \\
  & \iff
   \str G\models \bigvee_{0\le \ell\le k}
    \Big(\chi^{\ell}_{\chi_{\varphi}^{k+1}}
     \wedge \neg\chi^{\ell+1}_{\chi_{\varphi}^{k+1}}
     \wedge
     \bigvee_{0\le j\le (k-\ell)\cdot (k+1)}(\chi_{\textit{uni}}^{j}\wedge \neg \chi_{\textit{uni}}^{j+1}\wedge \rho_j)\Big).
\end{align}
Here the formula $\rho_j$, a formula expressing (in $\str G$ with a $\str
G'$ with exactly $j$ vertices) that $\str G'$ has a vertex cover of size
$k-\ell$, still has to be defined. We do that by saying that $\str G'$ (with
built-in arithmetic) is isomorphic to one of the graphs with $j$ vertices
(and with built-in arithmetics) that have vertex covers of size $k- \ell$.
For this we have to be able to define an order of $\str G'$ by a formula of
quantifier rank bounded by a constant number independent of $k$. Again this
is done with the color-coding method: We find $p$ and $q$, and $0\le
i_0<\cdots<i_{j-1} <j^2$ with
\[
h_{p,q}(G')= \{i_0,\ldots, i_{j-1}\}.
\]
Then, we can speak of the first, the second ,\ldots, vertex in $\str G'$.

As $\qr(\chi_{\varphi}^{k+1})\le 12$, we have $\qr(\textit{uni}(x))\le 13$.
Thus, $\qr(\chi_{\textit{uni}}^{j})\le 16$. As the remaining formulas
in~\eqref{eq:tab} have at most quantifier rank $16$, we get $\pVC\in
\X\FO_{16}$.
\proofend

\section{The hitting set problems with bounded hyperedge size}\label{sec:phs}

We consider the parameterized problem
\npprob{$\pdHS$}{A hypergraph $\str G$ with edges of size at most $d$}{
$k\in\mathbb N$}{Does $\str G$ have a hitting set of size $k$?}
A \emph{hypergraph} $\str G$ is a pair $(V,E)$, where $V$ is a set, the
set of \emph{vertices of $\str G$}, and every element of~$E$ is a
\emph{hyperedge}, that is, a nonempty subset of $V$. A \emph{hitting set} in
$\str G$ is a set $H$ that intersects each hyperedge (that is, $H\cap
e\ne \emptyset$ for all $e\in E$).

We view a hypergraph $\str G:= (V,E)$ as an $\{E_0, \epsilon\}$-structure
$\big(V\cup E, E,\epsilon^{\str G}\big)$, where $E_0$ is a unary relation
symbol and $\epsilon$ is a binary relation symbol and
\begin{eqnarray*}
E_0^{\str G}:= E
 & \text{and} &
\epsilon^{\str G}:=\big\{(v,e) \bigmid \text{$v\in V$, $e\in E$ and $v\in e$}\big\}.
\end{eqnarray*}
The goal of this section is to show:
\begin{theo}\label{thm:hit}
Let $d\ge 1$. Then $\pdHS$ is slicewise definable in $\FO$ with bounded
quantifier rank; more precisely, $\textsc{$p$-$d$-Hitting-Set}\in \FO_q $
with $ q=O(d^2)$.
\end{theo}

The following lemma can be viewed as a generalization of part of Buss'
kernelization algorithm for $\pVC$ to $\pdHS$. The case for
\textsc{$p$-$3$-Hitting-Set} was first shown in~\cite{nieros03}.

\begin{lem}\label{lem:hitset}
Let $(\str G,k)$ with $\str G=(V,E)$ be an instance of $\pdHS$. Let
$1<\ell\le d$ and assume that every $\ell$-set (i.e., set with exactly
$\ell$ elements) of vertices has at most $k^{d-\ell}$ extensions in $E$.

If $v_1,\ldots,v_{\ell-1}$ are pairwise distinct vertices such that there is
a hitting set $H$ of size $\le k$ that contains none of these vertices, then
$\{v_1,\ldots,v_{\ell-1} \}$ has at most $k^{d-(\ell-1)}$ extensions in~$E$.
\end{lem}

\proof Every hyperedge that extends $\{v_1,\ldots,v_{\ell-1}\}$ must contain
a vertex $u$ of the hitting set~$H$. By the assumptions, $u$ is distinct
from the $v_i$'s and therefore, the set $\{ v_1,\ldots,v_{\ell-1},u\}$ has
at most $k^{d-\ell}$ extensions in $E$. As $|H|\le k$, we see that there are
at most $k\cdot k^{d-\ell}\ \left(= k^{d-(\ell-1)}\right)$ extensions in
$E$. \proofend

Let $(\str G,k)$ and $1<\ell\le d$ satisfy the hypotheses of the lemma, that
is, $(\str G,k)$ with $\str G= (V,E)$ is an instance of $\pdHS$ and every
$\ell$-set has at most $k^{d-\ell}$ extensions in~$E$. For every pairwise
distinct vertices $v_1, \ldots, v_{\ell-1}$ such that $\{v_1, \ldots,
v_{\ell-1}\}$ has more than $k^{d-(\ell-1)}$ extensions in~$E$, we delete
from $E$ all hyperedges extending $\{v_1, \ldots, v_{\ell-1} \}$ and add the
hyperedge $\{v_1, \ldots, v_{\ell-1}\}$. Let $\str{G^{\ell}}= (V, E^{\ell})$
be the the resulting hypergraph. Then:
\begin{itemize}
\item[(a)] For every pairwise distinct vertices $v_1, \ldots, v_{\ell-1}$
    there are at most $k^{d-(\ell-1)}$ hyperedges in $E^{\ell}$ extending
    $\{v_1, \ldots, v_{\ell-1}\}$.

\item[(b)] If $H$ is a subset of $V$ and $|H|\le k$, then
    \begin{eqnarray*}
    \text{$H$ is a hitting set of $\str G$}
     & \iff &
    \text{$H$ is a hitting set of $\str G^{\ell}$},
    \end{eqnarray*}
    in particular,
    \begin{eqnarray*}
    (\str G,k)\in \pdHS
     & \iff &
    (\str G^{\ell},k)\in \pdHS.
    \end{eqnarray*}
\end{itemize}
Let $(\str G,k)$ be an instance of $\pdHS$. For $\ell:= d$ the hypothesis of
Lemma~\ref{lem:hitset} is fulfilled: Every $d$-set of vertices has at most
one extension in $E$, namely at most, itself. Hence, applying the above
procedure for $\ell= d$ we get the hypergraph $\str{G^{\ell}}$, which
satisfies the hypotheses of Lemma~\ref{lem:hitset} for $\ell:=d-1$. So we
get, again by the above procedure the hypergraph
$(\str{G^{\ell}})^{\ell-1}$, which we denote by~$\str{G^{\ell,\ell-1}}$.
Following this way, we finally obtain the hypergraph $\str{G^{\ell, \ell-1,
\ldots, 2}}$, which we denote by $\str G'$. Note that $\str G'= (V,E')$ for
some $E'$. From (a) and (b) we get (a$'$) and (b$'$).
\begin{itemize}
\item[(a$'$)] For every vertex $v$ there are at most $k^{d-1}$ hyperedges
    in $E'$ containing $v$.

\item[(b$'$)] If $H$ is a subset of $V$ and $|H|\le k$, then
    \begin{eqnarray*}
    \text{$H$ is a hitting set of $\str G$}
     & \iff &
    \text{$H$ is a hitting set of $\str G'$},
    \end{eqnarray*}
\end{itemize}
Moreover,
\begin{itemize}
\item[(c$'$)]\label{cprime} If $(\str G,k)\in
    \textsc{$p$-$d$-Hitting-Set}$, then $|E'|\le k^d$ and $|V'|\le d\cdot
    k^d$, where
    \[
    V':=\big\{v\in V \bigmid \text{there is an $e\in E'$ with
    $v$ in $e$}\}
    \]
    is the set of non-isolated vertices of $\str G'$.
\end{itemize}
In fact, let $H$ be a hitting set with $|H|= k$ of $\str G$ and hence, by
(b$ '$) of $\str G'$. As every hyperedge must contain a vertex of $H$, we
get $|E'|\le k^d$ from (a$'$). As every hyperedge $e\in E'$ contains at most
$d$ vertices, we have $|V'|\le d\cdot k^d$.

We fix $k$ and look at the $k$th slice of $\pdHS$. In the proof of
Theorem~\ref{thm:hit} we will see that for hypergraphs $\str G$ sufficiently
large compared with $k$ we can \FO-define $\str G'$ in $\str G$. By~(b$'$)
and~(c$'$), we know that $(\str G,k)\in \pdHS$ implies $|E'|\le k^d$. By
Theorem~\ref{thm:ccc}, we can express $|E'|\le k^d$ in first-order logic
with a bounded number of quantifiers if we add built-in addition and
multiplication. Essentially this shows that $\pdHS$ is eventually slicewise
definable in $\FO$ with bounded quantifier rank and thus, $\pdHS\in
\X\FO_{\qr}$ (by Proposition~\ref{pro:evesli}). This idea underlies the
following proof of Theorem~\ref{thm:hit}.
\medskip

\noindent \textit{Proof of Theorem~\ref{thm:hit}:}\label{app:hit} To
simplify the presentation we restrict ourselves to the case $d=3$.

Let $\str G_0= (V_0\cup E_0, E_0,\epsilon)$ be a hypergraph with hyperedges
of size at most three. Assume that $V_0:= \{1, \ldots, n\}$.




To present the application of the color-coding method in a readable fashion
we pass to a further structure $\str H$. Let $\sigma$ be the vocabulary
$\{\textit{Zero}, E, \textit{First}, \textit{Second}, \textit{Third}, < \}$,
where $\textit{Zero}$ and $E$ are unary relation symbols and all others
symbols are binary. Let $\str H$ be the $\sigma$-structure with
\begin{itemize}
\item $H= V^3$, the set of ordered triples of elements of $V:= V_0\cup
    \{0\}= \{0, 1, \ldots, n\}$\quad (for technical reasons, in $V$ we add
    $0$ to $V_0$),

\item $<^{\str H}$ is the lexicographic order on $V^3$,

\item $\textit{Zero}^{\str H}= \big\{(0,0,0)\big\}$,

\item $\textit{First}^{\str H}= \Big\{\big((u,v,w),(0,0,u)\big) \Bigmid
    (u,v,w)\in V^3\Big\}$,

\item $\textit{Second}^{\str H}= \Big\{\big((u,v,w),(0,0,v)\big) \Bigmid
    (u,v,w)\in V^3\Big\}$,

\item $\textit{Third}^{\str H}= \Big\{\big((u,v,w),(0,0,w)\big) \Bigmid
    (u,v,w)\in V^3\}$,

\item$ \begin{array}[t]{rcl}
E^{\str H}&=& \big\{(0,0,w)\bigmid \{w \}\in E_0\big\} \cup
    \big\{(0,v,w)\bigmid  \{v, w  \}\in E^{\str G}, \  v <w\big \}\ \cup\\[2mm]
&&\big\{(u,v,w)\bigmid \{u, v, w \}\in E^{\str G},  \ u< v< w\big\}.
\end{array}$

\end{itemize}
It is easy to show that $\str H$ can be defined by an \FO-interpretation in
$\str G$.

\medskip
We can express in $\str H$ that $x$ is an $i$-set (for $i= 1, 2, 3$) by an
\FO-formula $\varphi_{\textup{$i$-set}}$, where, say for $i=2$,
\[
\varphi_{\textup{$2$-set}}(x)
 := \exists x_1\exists x_2\exists x_3
  \Big(\textit{First}\, xx_1
   \wedge \textit{Zero}\,x_1
   \wedge \textit{Second}\, xx_2
   \wedge \textit{Third}\, xx_3
   \wedge x_1 <x_2<x_3\Big).
\]
Similarly, there is an \FO-formula $\varphi_{x\subseteq y}$ expressing that
``$x$ and $y$ are sets and that $x\subseteq y$.''

Fix $k\in \mathbb N$ and assume the vertex set $V^3$ of the hypergraph $\str
H$ is sufficiently large compared with~$k$. Furthermore, add built-in
addition and multiplication to $\str H$. Then we can \FO-define in $\str H$
the hypergraph corresponding to the hypergraph $\str H^3$ in the terminology
introduced after Lemma~\ref{lem:hitset}. In the transition to $\str H^3$ for
every 2-set~$ x$, which has more than $k$ extensions that are hyperedges, we
have to delete all these hyperedges and then add the hyperedge $x$. Note
that for the formula
\[
\varphi(x,y):= \big(\varphi_{x\subseteq y}(x,y)\wedge Ey\big)
\]
the \FO-formula $\chi_{\varphi}^{k+1}(x)$ expresses that ``$ x$ has more
than $k$ extensions that are hyperedges'' (see
Theorem~\ref{thm:ccc}). Thus, the new hyperedge relation (that is,
the hyperedge relation corresponding to the hypergraph $\str H^3$) is
given by
\begin{align*}
\varphi_{E^3}(x):=
 & \Big(\big(\varphi_{\textup{$1$-set}}(x)\wedge Ex\big)\ \vee \ \big(\varphi_{\textup{$ 2$-set}}(x)
 \wedge \chi_{\varphi}^{k+1}(x)\big)\\
 & \ \vee \
 \big(Ex \wedge \neg\exists y(\varphi_{\textup{$ 2$-set}}(y)\wedge \varphi_{x\subseteq y}(y,x)\wedge \neg\chi_{\varphi(y,z)}^{k+1}(y))\big)\Big).
\end{align*}
Similarly we can define the hyperedge relation corresponding to the
hypergraph $\str H^{3,2}=(V^3, E')$. By (b$'$) and (c$'$) on
page~\pageref{cprime}, we know that
\begin{quote}
$(\str H,k)\in \textsc{$p$-$3$-Hitting-Set} \iff (\str H^{3,2},k)\in
 \textsc{$p$-$3$-Hitting-Set}$, \\[1mm]
and if $(\str H,k)\in \textsc{$p$-$d$-Hitting-Set}$, then $|E'|\le k^3$
and $|V'|\le 3\cdot k^3$, where\\
$
V':=\big\{v\in V^3 \bigmid \text{there is an $e\in E'$ with $v$ in $e$}\}.
$
\end{quote}
So  the $k$th slice of $\pdHS$ can eventually be defined by a sentence
expressing
\[
 |E'|\le k^3~\text{and $((V', E'),k)$ is a \textsc{yes} instance of $\pdHS$.}
\]
Again such a formula is obtained using Theorem~\ref{thm:ccc} as in the proof
of Theorem~\ref{thm:pvc}.
\medskip

For the structures $\str G$ and \str H without built-in addition and
multiplication, we already saw that the second one can be obtained from the
first one by an \FO-interpretation. We need this result for the structures
with built-in addition and multiplication, too. This follows from
Proposition~\ref{pro:int}. Moreover, it is not hard to see that the
final \FO-sentence we obtain has quantifier rank $q = O(d^2)$.
\proofend

A part of an \FO-interpretation $I$ is an \FO-formula
$\varphi^I_{\textit{uni}}(x_1,\ldots, x_s)$ defining the universe of the
defined structure, that is: if $I$ is an interpretation of
$\sigma$-structures in a class~$\cls K$ of $\tau$-structures, then for every
structure $\str A\in \cls K$ the set
\[
(\varphi^{I}_{\textit{uni}})^{\str A}
 := \{(a_1,\ldots, a_s)\in A^s \mid \str A\models \varphi(a_1,\ldots, a_s)\}
\]
is the universe of the $\sigma$-structure $I(\str A)$ defined by $I$ in
$\str A$.

Assume that $\sigma$ does not contain the relation symbols $<,+,\times$, but
that the structures in~$\cls K$ are structures with built-in addition and
multiplication, i.e., $\cls K\se\ARITHM[\tau]$. In general, we can not extend the
interpretation $I$ to an interpretation $J$ such that
\[
J(\str A)= \left(I(\str A), <^{J(\str A) }, +^{J(\str A) }, \times^{J(\str A)}\right)
\]
has built-in addition and multiplication (that is, so that $J(\str A)$ is
$I(\str A)$ together with an order and the corresponding addition and
multiplication).

For example, for $\tau=\{P, <, +, \times\}$ with unary $P$ let $\cls K$ be the
class of $\tau$-structures \str A with $P^{\str A}\ne\emptyset$. Let
$\sigma$ be the empty vocabulary and consider the interpretation $I$
yielding in \str A the $\sigma$-structure with universe $P^{\str A}$ (take
$\varphi^I_{\textit{uni}}(x):= Px$). If we could extend $I$ to an
interpretation $J$ such that $J(\str A):=(P^{\str A}, <^{\str A}, +^{\str
A}, \times^{\str A})$ has built-in addition and multiplication, then we
could express in $J(\str A)$, and thus in $\str A$, that ``$P^{\str A}$ is
even,'' i.e., the parity problem, which is well known to be impossible.

The next result shows that the situation
is different if for $\varphi^I_{\textit{uni}}(x_1, \ldots, x_s)$ we have
$(\varphi^{I}_{\textit{uni}})^{\str A}= A^s$.

\begin{prop}\label{pro:int}
Let $\tau$ contain $<, +, \times$ and assume that none of these symbols is
in the vocabulary $\sigma$. Let $\cls K\subseteq \ARITHM[\tau]$ and let $I$ be an interpretation
of $\sigma$-structures in the structures in $\cls K$ with
$\varphi^{I}_{\textit{uni}}= \varphi^{I}_{\textit{uni}}(x_1,\ldots, x_s)$.
If for all $\str A\in \cls K$,
$$
(\varphi^{I}_{\textit{uni}})^{\str A}= A^s,
$$
then the interpretation $I$ can be extended to an interpretation of $\sigma\cup \{<, +, \times
\}$ such that $
J(\str A)= \big(I(\str A), <^{J(\str A)}, +^{J(\str A)}, \times^{J(\str A)}\big)
$
has built-in addition and multiplication for all $\str A\in \cls K$.
\end{prop}
\proof
Let $\str A\in \cls K$ and assume $A=\uni{n}$ and $<,+,\times$ have their
natural interpretations. We define the extension $J(\str A)$ of $I(\str A)$
(the construction will be independent of \str A). Of course, the
lexicographic order of $\uni{n}^s$ (the universe of $I(\str A)$) is
\FO-definable in \str A. So we define $J$ such that $<^{J(\str A)}$ is the
lexicographic order. Then $(a_1,\ldots, a_s)\in \uni{n}^s$ is the element at
the position
\[
a_1\cdot n^{s-1}+\cdots +a_{s-1}\cdot n+a_s
\]
in $<^{J(\str A)}$.

For $a,b\in A$ with $a+b\ge n$ and $0\le i<s $, we have
\[
a\cdot n^i+b\cdot n^i
 = n^{i+1} + (a+b-n)\cdot n^i
 = n^{i+1} + \big(a - (n-b)\big) \cdot n^i
\]
and $n-b$, $a - (n-b) \in A$. Thus, the built-in addition (with respect to
the lexicographic order) can be \FO-defined using $+^{\str A}$ by
formalizing the addition of base $n$ numbers with at most $s$ digits.
\medskip

The \FO-definition of the multiplication is not so easy. Note that
\[
\sum_{i=1}^{s}a_i\cdot n^{s-i}\cdot\sum_{j=1}^{s}b_j\cdot n^{s-j}
 = \sum_{k=0}^{2s-2}\left(\sum_{i+j=2s-k}a_ib_j\right)\cdot n^k.
\]
From this equation, we see that once we know how to \FO-define in $\str A$
the product $(0, \ldots, 0, a)\times (0, \ldots, 0, b)$ with $a, b \in A$,
we can \FO-define $(a_1, \ldots, a_s)\times (b_1, \ldots, b_s)$ for
arbitrary tuples in~$\uni n^s$. Of course, thereby taking into account
whether this product is $< n^s$. As $a\cdot b< n^2$ for $a,b\in \uni n$, we
can restrict ourselves to the case $s=2$, that is, we have to \FO-define
$(0,a)\times (0,b)$ with help of \FO-definition of the addition. We assume
$n> 2$ (and leave the case $n=2$ to the reader, the case $n=1$ being
trivial).

For this purpose we consider the smallest element $e\in \uni n$ such that
$e^2\ge n$ (\underline{e}xceeds $n-1$) and the \underline{l}argest element
$\ell\in \uni n$ such that $\ell^2\le n-1$. By $n>2$, we have
\begin{equation}\label{eq:exla}
e= \ell+1 \quad \text{and} \quad \ell+ \ell \le n-1
\end{equation}
and both, $e$ and $\ell$, are \FO-definable in $\str A$.

We first \FO-define $(0,e)\times(0,e)$. This will allow us to \FO-define
$(0,a)\times (0,b)$, essentially by writing $a$ and $b$ in base $e$
notation.

By~\eqref{eq:exla}, $e^2- \ell^2 = e + \ell$. Hence, $e^2=\ell +\ell+1
+\ell^2= n+ \ell + \ell - \big((n-1) - \ell^2\big)$. Thus,
\begin{equation}\label{eq:deft}
(0,e)\times (0,e)=(1,t)
 \quad \text{with}\quad
t = \ell + \ell - \big((n-1) - \ell^2\big).
\end{equation}
Note that $(n-1)- \ell^2 \in A$, thus by~\eqref{eq:exla}, $t\in A$.
Therefore we can \FO-define $(0,e)\times (0,e)$.

With the following two claims we will obtain the full result.
\medskip

\noindent \textit{Claim 1.} For $d\le \ell$ we can \FO-define $d\cdot e$,
$d\cdot t$, and $d\cdot e^2$.
\medskip

\noindent \textit{Proof of Claim 1:} $d\cdot e$: We have $d\cdot e= d\cdot
(\ell +1)= d\cdot \ell +d$. As $d\cdot \ell\le \ell^2\in A$, the claim
follows.

\medskip
\noindent $d\cdot t$: By~\eqref{eq:deft}, $t\le 2\cdot \ell$. Therefore
there is $t'\le \ell$ and $q\in \{0, 1\}$ with $t= t'+t'+q$. Hence, $d\cdot
t= d\cdot t'+ d\cdot t'+ d\cdot q$. As $d\cdot t'\in A$ and $d\cdot q\in
\{0,q\}$, the claim follows.

\medskip
\noindent $d\cdot e^2$: We know that $(0,e)\times (0,e)=(1,t)$ and
 $(0,d)\times (1,t)= (0,d)\times (1,0)+ (0,d)\times (0,t)$.
Clearly, $(0,d)\times (1,0)= (d,0)$. Furthermore, we know how to \FO-define
$d\cdot t$ by the previous step. Therefore, the claim follows.
\hfill$\dashv$

\medskip
The following result extends Claim 1.

\medskip
\noindent \textit{Claim 2.} For $d\le n-1$ we can \FO-define $d\cdot e$,
$d\cdot t$, and $d\cdot e^2$.
\medskip

\noindent \textit{Proof of Claim 2:} We write $d$ in the form $d= d_1\cdot e
+ d_2$ with $d_1, d_2 \le \ell$ (recall that $(\ell+1)\cdot e= e^2\ge n$).

\medskip
\noindent $d\cdot e$: We have $d\cdot e= d_1\cdot e^2 + d_2\cdot e$ and the
result follows by Claim~1.
\medskip

\noindent $d\cdot t$: By~\eqref{eq:deft}, $t\le 2\cdot \ell< 2\cdot e$.
Thus, there are $t_1\in \{0,1\}$ and $t_2\le \ell$ with $t= t_1\cdot e+
t_2$. Therefore
\[
d\cdot t=d_1\cdot t_1\cdot e^2+d_1\cdot t_2\cdot e+d_2\cdot t_1\cdot e+ d_2\cdot t_2.
\]
If $t_1\ne 0$, then $d_1\cdot t_1\cdot e^2= d_1\cdot e^2$. As $d_1\le \ell$,
this term is \FO-definable by Claim~1. As $d_1\cdot t_2$ and $d_2\cdot t_1$
are $\le n-1$, the corresponding terms are \FO-definable by the first part
of this claim.

\medskip
\noindent $d\cdot e^2$: Recall that $e^2= (1,t)$. We have $(0,d)\times
(1,t)= (0,d)\times(1,0)+(0,d)\times (0,t)$ and we just saw how to \FO-define
$d\cdot t$. \hfill $\dashv$

\medskip
Now we turn to the general case. Let $a,b\in A$. We may write $a=a_1\cdot
e+a_2$ and $b=b_1\cdot e+ b_2$ with $a_1,a_2,b_1,b_2\le \ell$. Thus,
\[
a\cdot b=a_1\cdot b_1\cdot e^2+a_1\cdot b_2\cdot e+a_2\cdot b_1\cdot e+ a_2\cdot b_2.
\]
As the products $a_1\cdot b_1$, $a_1\cdot b_2$, $a_2\cdot b_1$, and
$a_2\cdot b_2$ are all $\le n-1$, the result follows by Claim~2. \proofend

The following result, applied in Section~\ref{sec:paraac0}, extends
Proposition~\ref{pro:int} to interpretations whose universe are definable
initial segments of a Cartesian product.
\begin{cor}\label{cor:int}
Let $\tau$ contain $<, +, \times$ and assume that none of these symbols is
in the vocabulary $\sigma$. Let $\cls K\subseteq \ARITHM[\tau]$ and let $I$ be an interpretation
of $\sigma$-structures in the structures in $\cls K$ with
$\varphi^{I}_{\textit{uni}}= \varphi^{I}_{\textit{uni}}(x_1,\ldots, x_s)$.
Let $\cls K\subseteq \ARITHM[\tau]$ and let $I$ be an interpretation of
$\sigma$-structures in the structures in $\cls K$ with
$\varphi^{I}_{\textit{uni}}= \varphi^{I}_{\textit{uni}}(x_1,\ldots, x_s)$.
Furthermore, assume that there is an \FO-formula
$\varphi_{\textit{init}}(x_1, \ldots, x_s)$ such that for all $\str A\in \cls K$,
there is a unique tuple in $A^s$, we denote it by $(a_1,\ldots, a_s)$, such
that
\[
\str A\models \varphi_{\textit{init}}(a_1,\ldots, a_s)
 \quad \text{and} \quad
(\varphi^{I}_{\textit{uni}})^{\str A}
 = \big\{(b_1,\ldots, b_s)\in A^s \bigmid (b_1,\ldots, b_s)<_{\textup{lex}}(a_1,\ldots, a_s)\big\}
\]
\big(here $<_{\textup{lex}}$ denotes the lexicographic order with respect to
$<^{\str A}$\big). Then $I$ can be extended to an interpretation of
$\sigma\cup \{<, +, \times\}$ such that $J(\str A)= \left(I(\str A),
<^{J(\str A)}, +^{J(\str A)}, \times^{J(\str A)}\right)$ has built-in
addition and multiplication for all $\str A\in\cls K $.
\end{cor}

\section{Fagin definability}\label{sec:fagin}
Let $\varphi(X)$ be an $\FO[\tau]$-formula which for a, say $r$-ary,
second-order variable $X$ may contain atomic formulas of the form
$Xx_1\ldots x_r$ . Then the \emph{parameterized problem $\FD_{\varphi(X)}$
Fagin-defined by $\varphi(X)$} is the problem
\npprob{$\FD_{\varphi(X)}$}{A
$\tau$-structure \str A}{$ k\in \mathbb N$}{Decide whether there is an $S\se
A^r$ with $|S|=k$ and $\str A\models \varphi(S)$.}

The following metatheorem improves~\cite[Theorem~4.4]{flugro06}.

\begin{theo}\label{thm:fag}
Let $\varphi(X)$ be an $\FO[\tau]$-formula without first-order variables
occurring free and in which $X$ does not occur in the scope of an
existential quantifier or negation symbol. Then
$\FD_{\varphi(X)}\in\X\FO_{\qr}$ that is, $\FD_{\varphi(X)}$ is slicewise
definable with bounded quantifier rank.
\end{theo}

Recall that we view a hypergraph $\str G:= (V,E)$ as an $\{E_0, \epsilon\}$-structure
$\big(V\cup E, E,\epsilon^{\str G}\big)$, where $E_0$ is a unary relation
symbol and $\epsilon$ is a binary relation symbol and
\begin{eqnarray*}
E_0^{\str G}:= E
 & \text{and} &
\epsilon^{\str G}:=\big\{(v,e) \bigmid \text{$v\in V$, $e\in E$ and $v\in e$}\big\}.
\end{eqnarray*}
Fix $d\in\mathbb N$. For $k\in \mathbb N$ we have (assuming $|V|\ge k$)
\begin{eqnarray*}
(\str G,k)\in\pdHS &\iff &
\text{for some $S$ with $|S|=k$ we have $(V\cup E, E,\epsilon^{\str G})\models \varphi(S)$},
\end{eqnarray*}
where $\varphi(X):=\forall e\Big(E_0e\to \forall x_1\ldots \forall x_d
\big((\forall x (x\, \epsilon\, e\leftrightarrow \bigvee_{i=1}^d x_i=x)\to
(Xx_1\vee\ldots \vee Xx_d)\big)\Big)$.
By Theorem~\ref{thm:fag} we know that $\FD_{\varphi(X)}\in \X\FO_{\qr}$.
Hence, $\pdHS\in \X\FO_{\qr}$, so we get the result of the previous section.
However here, to prove Theorem~\ref{thm:fag} we use the result of the
previous section.
\medskip

\noindent
\textit{Proof of Theorem~\ref{thm:fag}:} For simplicity, let us
assume that $X$ is unary. Without loss of generality we can assume that
\[
\varphi(X)=\forall y_1\ldots\forall
y_\ell\bigwedge_{i=1}^m\bigvee_{j=1}^{p}\psi_{ij},
\]
where each
$\psi_{ij}$ either is $Xy_q$ for some $q\in\{1,\ldots,\ell\}$, or a
first-order formula with free variables in $\{y_1,\ldots,y_\ell\}$ in
which $X$ does not occur.

Let $(\str A,k)$ be an instance of $\FD_{\varphi(X)}$. We construct an
instance $(\str G(\str A),k)$ of $p\textsc{-$\ell$-Hitting-Set}$ such that
\begin{eqnarray}\label{eqn:fahi}
 (\str A,k)\in\FD_{\varphi(X)} & \iff &
(\str G(\str A),k)\in p\textsc{-$\ell$-Hitting-Set}.
\end{eqnarray}
As $(\str G(\str A),k)$ we take the hypergraph $(V,E)$ with $V=A$ and where
$E$ contains the following hyperedges. Let $\bar a\in A^{\ell}$ and $i\in
\{1,\ldots, m\}$. If
\[
\str A\models \neg\bigvee_{\substack {j\in \{1,\ldots, p\}\\
\text{ $X$ does not occur in $\psi_{ij}$ }}} \psi_{ij}(\bar a).
\]
then $E$ contains the hyperedge $\{a_{s_1},\ldots, a_{s_t} \}$ where
$Xy_{s_1},\ldots, Xy_{s_t}$ are exactly the disjuncts of the form
$Xy_{\ldots }$ in $\bigvee_{j=1}^{p}\psi_{ij} $. If $t=0$ (for some $\bar
a\in A^{\ell} $), we take as $\str G(\str A)$ a fixed hypergraph chosen in
advance such that $(\str G(\str A),k)$ is a \textsc{no} instance of
$p\textsc{-$\ell$-Hitting-Set} $.

Since $\str G(\str A)$ can be defined from $\str A$ by an \FO-interpretation
and $p\textsc{-$\ell$-Hitting-Set}\in \X\FO_{\qr}$, we get
$\FD_{\varphi(X)}\in \X\FO_{\qr}$.
\proofend

Some parameterized problems can be shown to be in $\para\FO$ by a simple
application of this theorem, e.g., for every $\ell\ge 1$, the problem
$p\textsc{-WSat}(\Gamma^+_{1,\ell})$, the restriction of
$p\textsc{-Dominating-Set}$ to graphs of degree $\ell$, and the problem
$p\textsc{-$\ell$-Matrix-Domination}$. Let us consider one example in detail
\npprob{$p\textsc{-$\ell$-Matrix-Domination}$}{An $n\times n$ matrix $M$
with entries from $\{0, 1\}$, which has in every row and in every column at
most $\ell$ ones and $k\in \mathbb N$}{$k$}{Is there a set $S$ of $k$
nonzero entries in $M$ that dominate all others, in the sense that every
nonzero entry in $M$ is in the same row or in the same column as some
element of $S$?}
We assign to such a matrix $M$ the structure $\str A(M):=(\uni n,
\textit{One}^{\str A(M)})$, where $\textit{One}^{\str A(M)}$, the
interpretation of the binary relation symbol $\textit{One}$, is
\[
\textit{One}^{\str A(M)}
 =\big\{(i,j)\in \uni n\times \uni n
  \bigmid \text{the $(i,j)$th entry of $M$ is 1} \big\}.
\]
Then for instances $(M,k)$ (with $|\textit{One}^{\str A(M)}|\ge k$), we have
\[
(M,k)\in p\textsc{-$\ell$-Matrix-Domination }\iff \str A(M)\in\FD_{\varphi(X)},
\]
where $\varphi(X)$ with binary $X$ is the following formula:

\begin{align*}
\forall x \forall y
 \Big( & \textit{One}xy\to
  \forall y_1\ldots \forall y_{\ell}\forall x_1\ldots \forall x_{\ell}
   \Big(\forall z(\textit{One} xz\leftrightarrow \bigvee_{1\le i\le \ell} z=y_i)\wedge \\
 & {\hspace{1cm}}(\forall z(\textit{One} zy\leftrightarrow
 \bigvee_{1\le i\le \ell} z=x_i))\Big)\to \bigvee_{1\le i\le \ell}(Xxy_i\vee Xx_iy)\Big).
\end{align*}

\section{$\para\AC^0=\X\FO_{\qr}$}\label{sec:paraac0}
The importance of the class $\X\FO_{\qr}$ from the point of view of
complexity theory stems from the fact that it coincides with the class
$\para\AC^0$, the class of parameterized problems that are in
dlogtime-uniform $\AC^0$ after a precomputation. As dlogtime-uniform $\AC^0$
contains precisely the class of parameterized problems definable in
first-order logic, the class $\para\AC^0$ corresponds to the class
$\para\FO$ of parameterized problems definable in first-order logic after a
precomputation on the parameter (see~\cite{elbstotan15,cheflu16}). We deal
here with the class $\para\FO$ and thus in this section aim to show
$\para\FO= \X\FO_{\qr}$.
\medskip

To define the class $\para\FO$ we need a notion of union of two arithmetical
structures.
\begin{defn}\label{def:oplus}
Assume $\str A\in \ARITHM[\tau]$ and $\str A'\in \ARITHM[\tau']$ satisfy
\begin{eqnarray*}
A\cap A'= \emptyset & \text{and} & \tau\cap \tau'= \{<, +, \times\}.
\end{eqnarray*}
Let $U$ be a new unary relation symbol. We set $\tau\uplus\tau' := \tau\cup
\tau'\cup \{U\}$. Then $\str A\uplus \str A'$ is the structure $\str{B}\in
\ARITHM(\tau\uplus\tau')$ with
\begin{itemize}
\item $B:=A\cup A'$;

\item $U^{\str{B}}=A'$;

\item $<^{\str{B}}:= <^{\str A}\cup <^{\str A'}\cup \big\{(a,a') \bigmid
    \text{$a\in A$ and $a'\in A'$}\big\}$, that is, the order
    $<^{\str{B}}$ extends the orders $<^{\str A}$ and $<^{\str A'}$, and
    in $<^{\str{B}}$ every element of $A$ precedes every element of $A'$;

\item $R^{\str{B}}:= R^{\str A}$ for $R\in \tau$ and $R^{\str{B}}:=
    R^{\str A'}$ for $R\in \tau'$.
\end{itemize}
If $A\cap A'\ne \emptyset$, then we pass to isomorphic structures with
disjoint universes before defining $\str A\uplus \str A'$.
\end{defn}

\begin{defn}\label{def:grun}
Let $Q\subseteq \ARITHM[\tau]\times \mathbb N$ be a parameterized problem.
$Q$ is \textit{first-order definable after a precomputation}, in symbols
$Q\in \para\FO$, if for some vocabulary $\tau'$ there is a computable
function $\textit{pre}: \mathbb N \to \ARITHM[\tau']$, a
\textit{precomputation}, and a sentence $\varphi\in
\FO\big[\tau\uplus\tau'\big]$ such that for all $(\str A, k)\in
\ARITHM[\tau]\times \mathbb N$,
\[
(\str A, k)\in Q \iff \str A\uplus \textit{pre}(k)\models \varphi.
\]
\end{defn}

\noindent The main result of this section reads as follows. It is the
modeltheoretic analogue of the equivalence between (i) and (ii)
of~\cite[Proposition~6]{cheflu16}.\footnote{Proposition~6 in~\cite{cheflu16}
contains a third statement equivalent to (i) and (ii). The corresponding
modeltheoretic analogue \emph{decidable and eventually in \FO} also
characterizes $ \X\FO_{\qr}$.}
\begin{theo}\label{thm:ac0char}
$\para\FO= \X\FO_{\qr}$.
\end{theo}

In the proof we shall need the following lemma. Its proof uses the fact that
every computable function may be defined on the natural numbers (with
addition and multiplication) by a $\Sigma_1$-sentence (that is, by an
$\FO$-sentence of the form $\exists x_1\ldots \exists x_n \psi$ with
quantifier free $\psi$).

\begin{lem}\label{lem:nfxyg}
Let $f:\mathbb N\to \mathbb N$ be a computable function. Then there is an
$\FO\big[\{<,+,\times\}\big]$-formula $\psi_f(x,y)$ and an increasing and
computable function $g:\mathbb N\to \mathbb N$ with $g(m)>f(m)$ for
$m\in\mathbb N$ such that for all $n,a\in \mathbb N$ with $n\ge g(a)$ and
$b\in \uni n$,
\begin{eqnarray*}
\left(\uni n,<^{\uni n}, +^{\uni n}, \times^{\uni n}\right) \models \psi_f(a,b)
 & \iff &
f(a)= b.
\end{eqnarray*}
The obvious generalization of this result to functions $f:\mathbb N^s\to
\mathbb N$ for some $s\ge 1$ holds, too.
\end{lem}

\noindent
\textit{Proof of Theorem~\ref{thm:ac0char}:} Assume that $Q\in
\para\FO$. Hence, for some vocabulary $\tau'$ there is a computable function
$\textit{pre}: \mathbb N \to \ARITHM[\tau']$ and a sentence $\varphi\in
\FO[\tau\uplus\tau']$ such that for all $(\str A, k)\in \ARITHM[\tau]\times
\mathbb N$,
\[
(\str A, k)\in Q \iff \str A\uplus \textit{pre}(k)\models \varphi.
\]
Clearly, then $Q$ is decidable. Therefore, by Lemma~\ref{pro:evesli}, it
suffices to show that for some $q\in\mathbb N$ the problem $Q$ is eventually
slicewise definable in $\FO_q$, that is, that there are an increasing and
computable function $g:\mathbb N\to \mathbb N$ and computable functions
$k\mapsto m_k$ and $k\mapsto \psi_k\in \FO_q[\tau\cup C(m_k)]$ such that for
all $(\str A, k)\in \ARITHM[\tau]\times \mathbb N$ with $|A|\ge g(k)$ we
have
\begin{equation}\label{eq:eve2}
\str A\uplus \textit{pre}(k)\models \varphi\iff \str A_{C(m_k)}\models \psi_k.
\end{equation}
The main idea: As the precomputation $\textit{pre}$ is computable, for
$(\str A, k)\in \ARITHM[\tau]\times \mathbb N$ with sufficiently large $|A|$
compared with $|\textit{pre}(k)|$, we can \FO-define $\textit{pre}(k)$ in
$\str A_{C(k+1)}$. Furthermore, from $\str A$ and from this \FO-defined
$\textit{pre}(k)$ in $\str A_{C(k+1)}$ we get (an isomorphic copy of) $\str
A\uplus \textit{pre}(k)$ in $\str A_{C(k+1)}$ by an \FO-interpretation.
Summing up, we can \FO-interpret $\str A\uplus \textit{pre}(k)$ in $\str
A_{C(k+1)}$. This \FO-interpretation yields the desired $\psi_k$ satisfying
\eqref{eq:eve2}.

Some details: Let $\tau'$, the vocabulary of $\textit{pre}(k) $, be the set
$\{<,+,\times, R_1,\ldots, R_m\}$, where $R_i$ is of arity $r_i$. Recall
that $\textit{pre}$ is computable. Thus there is a computable function $f:
\mathbb N\to \mathbb N$ with
\[
f(k)=|\textit{pre}(k)|.
\]
We may assume that the universe of $\textit{pre}(k)$ is $\uni{f(k)}$ and
$<,+,\times$ have their natural interpretations in $\textit{pre}(k) $. For
easier presentation, let us assume that the same holds for $\str A$; so, in
particular, $\uni {|A|}$ is the universe of \str A.

For $i$ with $1\le i\le m$ let $h_i: \mathbb N^{1+r_i}\to \{0,1\}$ be the
computable function with
\[
h_i(k, b_1,\ldots, b_{r_i} )=1
 \iff
\left(\,b_1,\ldots, b_{r_i}<f(k) \quad \text{and} \quad R_i\,^{\textit{pre}(k) }b_1,\ldots, b_{r_i}\right).
\]
As $f$ and $h_1,\ldots, h_m$ are computable, \big(we know that they are
\FO-definable in arithmetic and\big) by Lemma~\ref{lem:nfxyg}, there is a
computable and increasing function $g:\mathbb N\to\mathbb N$ with $g(k)>
f(k)$ and there are \FO-formulas $\psi_f(x,y)$ and $\psi_{h_i}(x,y_1,\ldots,
y_{r_i})$ such that for the relevant arguments, the formulas $\psi_f(x,y)$
and $\psi_{h_i}(x,y_1,\ldots, y_{r_i})$ correctly define $f$ and $h_i$ in
models with built-in addition and multiplication of size $\ge g(k)$.
Clearly, once we have the values $f(k)$ and $h_i(k, b_1,\ldots, b_{r_i})$
for $1\le i\le m$ and $b_1,\ldots, b_{r_i}< f(k)$, we can first-order define
$\textit{pre}(k) $, and hence $(\str A_{C(k+1)}, R_1\,^{\textit{pre}(k)},
\ldots, R_m\,^{\textit{pre}(k)})$, in $\str A_{C(k+1)}$, whenever $|A|\ge
g(k)$.

By Corollary~\ref{cor:int} there is an \FO-interpretation yielding the
structure $\str A \uplus \textit{pre}(k)$ from the structure $(\str
A_{C(k+1)}, R_1\,^{\textit{pre}(k)}, \ldots, R_m\,^{\textit{pre}(k)})$.
Putting these interpretations together, we obtain an $\FO$-interpretation
yielding $\str A\uplus \textit{pre}(k)$ in $\str A_{C(k+1)}$ assuming
$|A|\ge g(k) $. Thus we obtain from $\varphi$ an \FO-sentence $\psi_k$
satisfying the equivalence~\eqref{eq:eve2}.

\medskip
Now assume that $Q\in \X\FO_{\qr}$. Then there
is a $q\in \mathbb N$ and computable functions $k\mapsto m_k$ with $m_k\in
\mathbb N$ and $k\mapsto \varphi_k$ with $\varphi_k\in \FO_q\big[\tau\cup
C(m_k)\big]$ such that for all $(\str A, k)\in \ARITHM[\tau]\times \mathbb
N$,
\[
(\str A, k)\in Q\iff \str A_{C(m_k)}\models \varphi_k.
\]
We have to find a precomputation $\textit{pre}: \mathbb N\to \ARITHM[\tau']$
and an $\FO[\tau\uplus\tau']$-sentence $\varphi$ such that for all $(\str A,
k)\in \ARITHM[\tau]\times \mathbb N$,
\begin{equation}\label{eq:drein}
\str A_{C(m_k)}\models \varphi_k\iff \str A\uplus \textit{pre}(k) \models \varphi.
\end{equation}
Essentially $\textit{pre}(k)$ is the parse tree of $\varphi_k$ and the
sentence $\varphi$ expresses that $\str A_{C(m_k)}$ satisfies the sentence
given by this parse tree, that is, the sentence $\varphi_k$.

We can assume that every sentence of quantifier rank $\le q$ (and thus,
every $\varphi_k$) has the variables among $x_1,\ldots, x_q$ and is written
as a disjunction of conjunctions of atomic formulas and of formulas starting
with a quantifier.

Let $p_k$ be the number of nodes of the parse tree of $\varphi_k$. The
structure $\textit{pre}(k)\in \ARITHM[\tau']$ has universe $\uni {\max\{p_k,
m_k\}}$. The binary relation symbol $E$ is interpreted by the edge relation
of the parse tree. Then, besides $E$, the vocabulary $\tau'$ among others,
will contain unary relations \textit{Exists}, \textit{Forall},
\textit{X$_1$},\ldots, \textit{X$_q$}, \textit{And}, \textit{Or}, and
\textit{Neg}. Furthermore, for every relational symbol $R\in\tau$ (for
simplicity, we consider a binary $R$) we need in $\tau'$ the unary relation
symbols
\[
\textit{At-}R, \ \textit{V11-}R,\ldots ,\textit{V1q-}R, \ \textit{V21-}R,\ldots,\textit{V2q-}R
\]
and the binary relation symbols
\[
\textit{C1-}R, \ \textit{C2-}R.
\]
For example, for a node $u$, for $1\le j\le q$, and for $i < m_k$ we have:
\begin{eqnarray*}
\textit{Exists}\ u
 & \iff & \text{ the node $u$ corresponds to an existentially quantified variable}\\
\textit{X$_j$}\ u
 & \iff & \text{ the quantifier in $u$ binds the variable $x_j$}\\
\textit{Or}\ u
 & \iff & \text{ $u$ corresponds to a disjunction}\\
\textit{At-}R\ u
 & \iff & \text{ $u$ corresponds to an atomic formula with the relation symbol $R$}\\
\textit{V1j-}R\ u
 & \iff & \text{ $u$ corresponds to an atomic formula of the form $Rx_j\ \cdot$}\\
\textit{V2j-}R\ u
 & \iff & \text{ $u$ corresponds to an atomic formula of the form $R \cdot \ x_j$}\\
\textit{C2-}R\ u\ i
 & \iff & \text{ $u$ corresponds to an atomic formula of the form $R \cdot \ \overline i$.}\\
\end{eqnarray*}
We leave it to the reader to write down a sentence $\varphi$
satisfying~\eqref{eq:drein}.
\proofend

\begin{cor}
For every $d\in\mathbb N$, $\pdHS$ is in $\para\FO$ (and hence in
$\para\AC^0$).
\end{cor}

\section{The hierarchy $(\FO_q)_{q\in\mathbb N}$ on arithmetical
structures}\label{sec:FOqhier}

Let $\tau_0:= \{<, +, \times\}$ and let $\tau$ be a vocabulary with
$\tau_0\subseteq \tau$. For $q\in \mathbb N$ by  $\FO_{q}[\tau]\subsetneq
\FO_{q+1}[\tau]$ \emph{on arithmetical structures} we mean that there is an
$\FO_{q+1}[\tau]$-sentence which is not equivalent to any
$\FO_{q}[\tau]$-sentence on all finite $\tau$-structures with built-in
addition and multiplication. We say that \emph{the hierarchy
$\big(\FO_q\big)_{q\in \mathbb N}$ is strict on arithmetical structures} if
there is a vocabulary $\tau\supseteq \tau_0$ such that $\FO_{q}[\tau]
\subsetneq \FO_{q+1}[\tau]$ on arithmetical structures for every $q\in
\mathbb N$.

\begin{theo}\label{thm:sip}
The hierarchy $\big(\FO_q\big)_{q\in\mathbb N}$ is strict on arithmetical
structures.
\end{theo}

Some preparations are in order. First, we recall how structures are
represented by strings. Let $\tau$ be a relational vocabulary and $n\in
\mathbb N$. We encode a $\tau$-structure $\str A$ with $A=[n]$  by a binary
string $\enc(\str A)$ of length
\[
\ell_{\tau,n}:= \sum_{R\in \tau} n^{\arity(R)}.
\]
For instance, assume $\tau= \{E,P\}$ with binary $E$ and unary $P$, then
\[
\enc(A)= i_0 i_1\cdots i_{n^2-1}\; j_0 j_1\cdots j_{n-1},
\]
where for every $a,b\in [n]$, $\big(i_{a+b\cdot n} = 1 \iff (a,b)\in E^{\str
A}\big)$ and $ (j_a=1 \iff a\in P^{\str A})$.
\begin{eqnarray*}
i_{a+b\cdot n} = 1 & \iff & (a,b)\in E^{\str A}, \\
j_a=1 & \iff & a\in P^{\str A}.
\end{eqnarray*}
Let $\cls K$ be a class of $\tau$-structures. A family of
circuits $(\C_n)_{n\in \mathbb N}$ \emph{decides} $\cls K$ if
\begin{enumerate}
\item every $\C_n$ has $\ell_{\tau,n}$ inputs,

\item for  $n\in \mathbb N$ and every $\tau$-structure $\str A$ with
    $A=[n]$, \ $( \str A\in \cls K  \iff  \C_n(\enc(\str A))=1) $.
\end{enumerate}

Recall that for $n\in \mathbb N$ the classes $\Sigma_n$ and $\Pi_n$ of
formulas are defined as follows: $\Sigma_0$ and $\Pi_0$ are the class of
quantifier free formulas. The class $\Sigma_{n+1}$ (the class $\Pi_{n+1}$)
is the class of formulas of the form $\exists x_1\ldots \exists x_k \varphi$
with $\varphi\in\Pi_n$ and arbitrary $k$ (of the form $\forall x_1\ldots
\forall x_k \varphi$ with $\varphi\in\Sigma_n$ and arbitrary $k$).

\begin{lem}\label{lem:qrpn}
Every \FO-formula of quantifier rank $q$ is logically equivalent to a
$\Sigma_{q+1}$-formula and to a $\Pi_{q+1}$-formula.
\end{lem}
\proof The proof is by induction on $q$. For $q=0$ the claim is trivial. The
induction step follows from the facts:
\begin{itemize}
\item An \FO-formula of quantifier rank $q+1$ is a Boolean combination of
    formulas of the form $\exists x\psi$ and $\forall x\psi$, where $\psi$
    has quantifier rank $\le q$. In formulas of the form $\exists x\psi$
    we replace, using the induction hypothesis, the formula $\psi$ by an
    equivalent $\Sigma_{q+1}$-formula, in formulas of the form $\forall x
    \psi$ we replace the formula $\psi$ by an equivalent
    $\Pi_{q+1}$-formula.

\item Boolean combinations of $\Sigma_{q+1}$-formulas and of
    $\Pi_{q+1}$-formulas are equivalent to both, a $\Sigma_{q+2}$-formula
    and to a $\Pi_{q+2}$-formula.
\end{itemize}\proofend

\begin{lem}\label{lem:FOq}
Let $q\in \mathbb N$. Then for every sentence $\varphi\in \FO_q$ there is a
family of circuits $\big(\C_n\big)_{n\in \mathbb N}$ of depth $\le q+2$ and
size $n^{O(1)}$ which decides $\Mod(\varphi)= \big\{\str A\mid \str A\models
\varphi\big\}$. Moreover, the output of $\C_n$ is an \textup{OR} gate, and
the bottom layer of gates in $\C_n$ has fan-in bounded by a constant which
only depends on $\varphi$.
\end{lem}

\proof
To simplify the discussion, we assume $q=3$. The other cases can be
proved along the same lines. By Lemma~\ref{lem:qrpn} the sentence $\varphi$
is equivalent to a $\Sigma_4$-sentence
\[
\psi = \exists x_{1,1}\cdots \exists x_{1,i_1} \forall x_{2,1} \cdots \forall x_{2,i_2}
\exists x_{3,1}\cdots \exists x_{3,i_3} \forall x_{4,1} \cdots \forall x_{4,i_4}
 \bigwedge_{p\in I_{\wedge}}\bigvee_{q\in I_{\vee}} \chi_{pq},
\]
where $I_{\wedge}$ and $I_{\vee}$ are index sets and every $\chi_{pq}$
is a literal.

For $n\in \mathbb N$ we construct the desired circuit $\C= \C_n$  using the
standard translation from \FO-sentences to $\AC^0$-circuits. That is, every
existential (universal) quantifier corresponds to a $\bigvee$ ($\bigwedge$)
gate with fan-in $n$; the conjunction is translated to a $\bigwedge$ gate
with fan-in $|I_{\wedge}|$ and the disjunctions to $\bigvee$ gates with
fan-in $|I_{\vee}|$. Next we merge consecutive layers of gates that are all
$\bigwedge$, or that are all $\bigvee$. The resulting circuit $\C_n$ is of
depth $q+2$. It has an \textup{OR} as output gate and bottom fan-in bounded
by $|I_{\vee}|$.
{\hspace*{3cm}}\proofend

Key to our proof of Theorem~\ref{thm:sip} are the following Boolean
functions, also known as \emph{Sipser functions}.
\begin{defn}[\cite{sip83,bopsip90}]\label{defn:sipser}
Let $d\ge 1$  and $m_1, \ldots, m_d\in \mathbb N$. For every $i_1\in [m_1]$,
$i_2\in [m_2]$, \ldots, $i_d\in [m_d]$ we introduce a Boolean variable
$X_{i_1, \ldots, i_d}$. Define
\begin{equation}\label{eq:sipser}
f^{m_1, \ldots, m_d}_d := \bigwedge_{i_1\in [m_1]} \bigvee_{i_2\in [m_2]}
\cdots \bigodot_{i_d\in [m_d]} X_{i_1, \ldots, i_d},
\end{equation}
where $\bigodot$ is $\bigvee$ if $d$ is even, and $\bigvee$ otherwise. For
every $d\ge 2$ and $m\ge 1$ we set
\[
\sipserf^m_d := f^{m_1, \ldots, m_d}_d
\]
with $m_1= \ceil{\sqrt{m/\log\; m}}$, $m_2=\cdots m_{d-1}= m$, and $m_d=
\ceil{\sqrt{d/2\cdot m\cdot \log\; m}}$.

Observe that the size of $\sipserf^m_d$ is bounded by $m^{O(d)}$.
\end{defn}

The following lower bound for $\sipserf^m_d$ is proved in~\cite{has89}. We
use the version presented as Theorem 4.2 in~\cite{segbusimp04}.

\begin{theo}\label{thm:siplb}
Let $d\ge 2$. Then there exists a constant $\beta_d> 0$ so that if a depth
$d+1$, bottom fan-in $k$ circuit with an \textup{OR} gate as the output and
at most $S$ gates in levels $1$ through $d$ computes $\sipserf^m_d$, then
either $S\ge 2^{m^{\beta_d}}$ or $k\ge m^{\beta_d}$.
\end{theo}

\noindent
\textit{Proof of Theorem~\ref{thm:sip}:} $\FO_0 \subsetneq \FO_1$ is
trivial by considering the sentence $\exists x\; Ux$ where $U$ is a unary
relation symbol.  We still need to show that for an appropriate vocabulary
$\tau\supseteq \tau_0$ it holds $\FO_q[\tau]\subsetneq \FO_{q+1}[\tau]$ on
arithmetical structures for every $q\ge 1$.

Let $d, m\in\mathbb N$. We identify the function $\sipserf^m_d$ with the
circuit in~\eqref{eq:sipser} which computes it. Let $E$ be a binary relation
symbol and $U$ a unary relation symbol. Then we view the underlying
(directed) graph of $\sipserf^m_d$ as a $\{E,U\}$-structure $\str A_{d,m}$
with
\begin{align*}
A_{d,m} &:= \big\{v_g \bigmid \text{$g$ a gate in $\sipserf^m_d$}\big\}, \\
E^{\str A_{d,m}} &:= \big\{(v_{g'}, v_{g}) \bigmid \text{$g'$ is an input to $g$}\big\}, \\
U^{\str A_{d,m}} &:= \big\{v_g\bigmid \text{$g$ is an input to the output gate}\big\}.
\end{align*}
Let $P$ be a unary relation symbol. Every assignment $B$ of (truth values to
the input nodes of) $\sipserf^m_d$ can be identified with $P^{\str
A_{d,m}}:= \{g\mid \text{$g$ an input gate assigned to \true\ by $B$}\}$.
For $\tau':=\{E,U,P)$ we define an $\FO[\tau']$-sentence $\varphi_d$  such
that for all $m$,
\begin{eqnarray}\label{eq:psid}
\sipserf^m_d(P^{\str A_{d,m}})
 = \true & \iff &
  (\str A_{d,m}, P^{\str A_{d,m}})
 \models \varphi_d.
\end{eqnarray}
Fix $q\ge 1$. Assume $q$ is even and set $d:=q+1$ (the case of odd $q$ is
treated similarly). We define inductively $\FO[\tau']$-formulas
$\psi_\ell(x)$ by
\begin{align*}
\psi_0(x) := Px, \qquad \text{ and}\qquad \psi_{\ell+1}(x) & :=
 \begin{cases}
  \forall y \big(Eyx \to \psi_\ell(y)\big) & \text{if $\ell$ is even}, \\
  \exists y\big(Eyx \wedge \psi_\ell(y)\big) & \text{if $\ell$ is odd}.
 \end{cases}
\end{align*}
We set \big(recall the definition of $U^{\str A_{d,m}}$\big)
\begin{align*}
\varphi_{q+1} := \forall x (Ux \to \psi_{q}(x)).
\end{align*}
It is straightforward to verify that $\qr(\varphi_{q+1})=q+1$ and that
$\varphi_{q+1}$ satisfies \eqref{eq:psid} (for $ d=q+1$).

Let $\tau:= \tau' \cup\{<,+,\times\}= \{E,U,P,<,+,\times\}$. We define
\[
\sipser_{q+1} := \left\{\str A \in \ARITHM[\tau]
 \bigmid \str A\models \varphi_{q+1} \right\}.
\]
By definition the class $\sipser_{q+1}$ is axiomatizable in
$\FO_{q+1}[\tau]$. We show that $\sipser_{q+1}$ is not axiomatizable in
$\FO_{q}[\tau]$. For a contradiction, assume that $\sipser_{q+1}=
\Mod(\varphi)$ for some $\varphi\in \FO_{q}[\tau]$. Then by
Lemma~\ref{lem:FOq} there exists a family of circuits $\big(\C_n\big)_{n\in
\mathbb N}$ such that the following conditions are satisfied.
\begin{enumerate}
\item[(C1)] Every $\C_n$ has $\ell_{\tau,n}$ inputs, depth $q+2$, and size
    $\ell_{\tau,n}^{O(1)}$.

\item[(C2)] The output of $\C_n$ is an \textup{OR} gate, and its bottom
    fan-in is bounded by a constant.

\item[(C3)] For every $n\in \mathbb N$ and every $\tau$-structure $\str A$
    with $A=[n]$
    \begin{eqnarray*}
    \str A\in \sipser_{q+1} & \iff & \C_n(\enc(\str A))=1.
    \end{eqnarray*}
\end{enumerate}
Let $m\in \mathbb N$ and let $n$ be the number of variables in
$\sipserf^m_{q+1}$, i.e.,
\[
n= \ceil{\sqrt{m/\log m}}\cdot m^{q-1}
 \cdot \ceil{\sqrt{(q+1)/2\cdot m\cdot \log m}}.
\]
Consider the structure $\str A_{q+1,m}$ associated with $\sipserf^m_{q+1}$
and expand it with $<, +, \times$. Thus for any assignment of the $n$
inputs, identified with the unary relation $P^{\str A_{q+1,m}}$, we have
\begin{eqnarray*}
\sipserf^m_{q+1}(P^{\str A_{q+1,m}})=1 & \iff &
 \left(\str A_{q+1,m}, <, +, \times, P^{\str A_{q+1,m}}\right)\models \varphi \\
 & \iff & \C_n\Big(\enc\left(\str A_{q+1,m}, <, +, \times, P^{\str A_{q+1,m}}\right)\Big)=1.
\end{eqnarray*}
Here is the crucial observation. In the  string $\enc(\str
A_{q+1,m}, <, +, \times, P^{\str A_{q+1,m}})$ only the last $n$ bits
depend on the assignment, that is, on $P^{\str A_{q+1,m}}$. These are
precisely the $n$ input bits for the $\sipserf^m_{q+1}$ function. Thus we
can simplify the circuit $\C_n$ by fixing the values of the first
$\ell_{\tau, n}-n$ inputs according to $(\str A_{q+1,m}, <, +, \times)$. Let
$\C^*_n$ be the resulting circuit. We have
\begin{eqnarray*}
\sipserf^m_{q+1}(P^{\str A_{q+1,m}})=1
 & \iff &
\C^*_n(P^{\str A_{q+1,m}})=1.
\end{eqnarray*}
By (C1), $\C^*_n$ has depth $q+2$ and size $n^{O(1)}$ (as $\ell_{\tau,n}=
n^{O(1)}$). By (C2) its output is an \textup{OR}~gate, and its bottom fan-in
is bounded by a constant. As $m\in \mathbb N$ is arbitrary, this clearly
contradicts Theorem~\ref{thm:siplb}. {\hspace*{3cm}} \proofend

\noindent
\textit{Proof of Theorem~\ref{thm:FOqhier}:} Let $q\in
\mathbb N$. By Theorem~\ref{thm:sip} we know that there is a vocabulary
$\tau$ and an $\FO_{q+1}[\tau]$-sentence $\varphi$ which is not equivalent
to any $\FO_{q}[\tau]$-sentences on arithmetical structures. We claim that
\[
Q:= \big\{(\str A,0) \bigmid \text{$\str A\in \ARITHM[\tau]$ and $\str A\models \varphi$}\big\}
\]
is not slicewise definable in $\FO_{q}$. As $Q$ is slicewise definable in
$\FO_{q+1}$, this would give us the desired separation.

Assume otherwise, then, by Definition~\ref{def:sli}, there is a constant
$m_0\in \mathbb N$ and a sentence $\psi$ in $\FO_q[\tau\cup C(m_0)]$ such
that for every $\str A\in \ARITHM[\tau]$
\begin{eqnarray*}
\str A\models \varphi & \iff & \str A_{C(m_0)}\models \psi.
\end{eqnarray*}
This does not give us a contradiction immediately, since $\psi$ might
contain constants in $C(m_0)$. But it is easy to see that
Lemma~\ref{lem:qrpn} and Lemma~\ref{lem:FOq} both survive in the presence of
constants. Thus almost the same proof of Theorem~\ref{thm:sip} shows that
$\psi\in \FO_q[\tau\cup C(m_0)]$ cannot exist. \proofend

\section{Conclusions}
We have shown that a few parameterized problems are slicewise definable in
first-order logic with bounded quantifier rank. In particular, the
$k$-vertex-cover problem, i.e., the $k$th slice of $\pVC$, is definable in
$\FO_{16}$ for every $ k\in \mathbb N$. One natural follow-up question is
whether this is optimal. Or can we show at least that $\pVC\notin \X\FO_2$?
Such a question is reminiscent of the recent quest for optimal algorithms
for natural polynomial time solvable problems (see e.g.,~\cite{bacind15}).
In our result $\pdHS\in \X\FO_{q}$ we have $q = O(d^2)$, and we conjecture
that there is no universal constant $q$ which works for every $\pdHS$. But
so far, we do not know how to prove such a result.

It turns out that the class $\X\FO_{\qr}$ coincides with the parameterized
circuit complexity class $\para\AC^0$ which has been intensively studied
in~\cite{banstotan15,cheflu16}. Similar to~\cite{banstotan15}, it seems that
all the non-trivial examples in $\X\FO_{\qr}$ require the color-coding
technique. It would be interesting to see whether other tools from
parameterized complexity can be used to show membership in $\X\FO_{\qr}$.

We have also established the strictness of $\big(\X\FO_{q}\big)_{q\in
\mathbb N}$ by proving that $\FO_{q}\subsetneq \FO_{q+1}$ on arithmetical
structures for every $q\in \mathbb N$. Our proof is built on a strict
$\AC^0$-hierarchy on Sipser functions. We conjecture that the sentence
\[
\exists x_1 \cdots \exists x_{q+1} \bigwedge_{1\le i< j\le q+1} E_{x_ix_j},
\]
which characterizes the existence of a ($q+1$)-clique, witnesses
$\FO_{q}\subsetneq \FO_{q+1}$ on graphs with built-in addition and
multiplication. Rossman~\cite{ros08} has shown that ($q+1$)-clique cannot be
expressed in arithmetical structures with $ \lfloor (q+1)/4\rfloor$
variables and hence not in $ \FO_{ \lfloor (q+1)/4\rfloor }$. This already
shows that the hierarchy $ (\FO_q)_{q\in\mathbb N}$ does not collapse.
\bibliographystyle{plain}
\bibliography{FOr}

\end{document}